\documentclass[traditabstract]{aa}

\usepackage{graphicx}
\usepackage{natbib}
\usepackage{ulem}
\usepackage{siunitx}
\usepackage{wasysym}
\usepackage{subfiles}

\usepackage{xcolor}

\newcommand{\wa}[0]{\mbox{WASP--12}}
\newcommand{\wab}[0]{\mbox{WASP--12\,b}}

\newcommand{\hp}[0]{\mbox{HAT--P--32}}
\newcommand{\hpb}[0]{\mbox{HAT--P--32}\,b}
\newcommand{\hei}[0]{\ion{He}{i}\,$\lambda$10833\,{\AA}}

\newcommand{\ha}[0]{\mbox{H$\alpha$}}
\newcommand{\nad}[0]{\mbox{\ion{Na}{i}~D}}
\newcommand{\carm}[0]{CARMENES}

\newcommand{\cairt}[0]{\ion{Ca}{ii}~IRT}
\renewcommand{\ha}[0]{H$\alpha$}

\newcommand{\ms}[0]{m\,s$^{-1}$}
\newcommand{\kms}[0]{km\,s$^{-1}$}
\newcommand{\ergcmcms}[0]{erg\,cm$^{-2}$\,s$^{-1}$}

\newcommand{\gs}[0]{g\,s$^{-1}$}
\newcommand{\tc}[1]{$T$$_{\rm #1}$}

\newcommand{\xmm}[0]{\textit{XMM-Newton}}

\newcommand{\sys}[0]{\texttt{SYSREM}}
\newcommand{\gaia}[0]{\textit{Gaia}}

\newcommand{\rx}[1]{\mbox{$R$$_{\rm #1}$}}
\newcommand{\rs}[0]{\rx{\star}}
\newcommand{\rprs}[0]{\mbox{\rx{p}\,\rs$^{-1}$} }
\newcommand{\rp}[0]{\rx{p}}
\newcommand{\rj}[0]{\rx{Jup}}

\newcommand{\mx}[1]{\mbox{$M$$_{\rm #1}$}}
\renewcommand{\mp}[0]{\mx{p}}

\newcommand{\rsun}[0]{\rx{\odot}}
\newcommand{\mj}[0]{\mx{Jup}}

\newcommand{\mlp}[1]{\textcolor{red}{#1}}
\renewcommand{\mlp}[1]{}

\def\rp{$R_{\rm P}$}

\usepackage{txfonts}

\DeclareMathAlphabet\mathbfcal{OMS}{cmsy}{b}{n}

\begin{document}

\title{The elusive atmosphere of \wab} 
\subtitle{High-resolution transmission spectroscopy with CARMENES}
\titlerunning{Transmission spectroscopy}
\authorrunning{S. Czesla et al.}

\author{S.~Czesla \inst{\ref{instTLS}} 
    \and M.~Lamp\'on \inst{\ref{instIAA}}
  \and D.~Cont \inst{\ref{instLMU}, \ref{instExzO}}
  \and F.~Lesjak\inst{\ref{instIAG}}
  \and J.~Orell-Miquel\inst{\ref{instIAC}, \ref{instLaguna}}
  \and J.~Sanz-Forcada \inst{\ref{instCAB}}
  \and E.~Nagel\inst{\ref{instIAG}}
  \and L.~Nortmann \inst{\ref{instIAG}}
  \and K.~Molaverdikhani\inst{\ref{instLMU},\ref{instMPIA},\ref{instLSW}}
  \and M.~L\'opez-Puertas \inst{\ref{instIAA}}
  \and F.~Yan \inst{\ref{instFei}}
  \and A.~Quirrenbach\inst{\ref{instLSW}}
  \and J.~A.~Caballero\inst{\ref{instCAB}}
  \and E.~Pall\'e\inst{\ref{instIAC}, \ref{instLaguna}}
  \and J.~Aceituno\inst{\ref{instCAHA}}
  \and P.~J.~Amado\inst{\ref{instIAA}}
  \and
  Th. Henning\inst{\ref{instMPIA}}
  \and S. Khalafinejad\inst{\ref{instLSW}}
  \and D. Montes\inst{\ref{instUCM}}
  \and A.~Reiners\inst{\ref{instIAG}}
  \and I.~Ribas\inst{\ref{instICE},\ref{instIEEC}}
  \and A. Schweitzer\inst{\ref{instHS}}
  }      

\institute{
        Th\"uringer Landessternwarte Tautenburg, Sternwarte 5, 07778 Tautenburg, Germany \label{instTLS}\\ 
        \email{sczesla@tls-tautenburg.de}
        \and
        Instituto de Astrof\'isica de Andaluc\'{i}a (IAA-CSIC), Glorieta de la Astronom\'ia s/n, 18008 Granada, Spain\label{instIAA}
        \and
        Universit\"ats-Sternwarte, Ludwig-Maximilians-Universit\"at M\"unchen, Scheinerstrasse 1, 81679 M\"unchen, Germany\label{instLMU} 
        \and
         Exzellenzcluster Origins, Boltzmannstrasse 2, 85748 Garching, Germany\label{instExzO}
        \and
        Institut f\"ur Astrophysik und Geophysik, Georg-August-Universit\"at G\"ottingen, Friedrich-Hund-Platz 1, 37077 G\"ottingen, Germany\label{instIAG} 
        \and
        Instituto de Astrof\'{\i}sica de Canarias, c/ V\'{\i}a L\'actea s/n, 38205 La Laguna, Tenerife, Spain\label{instIAC}
        \and
        Departamento de Astrof\'{\i}sica, Universidad de La Laguna, 38206 Tenerife, Spain\label{instLaguna}
        \and
        Centro de Astrobiolog\'ia (CSIC-INTA), ESAC, Camino bajo del castillo s/n, 28692 Villanueva de la Ca\~nada, Madrid, Spain\label{instCAB}
        \and
        Max-Planck-Institut f\"ur Astronomie, K\"onigstuhl 17, 69117 Heidelberg, Germany\label{instMPIA}
        \and
        Landessternwarte, Zentrum f\"ur Astronomie der Universit\"at Heidelberg, K\"onigstuhl 12, 69117 Heidelberg,
        Germany\label{instLSW}
        \and
        Department of Astronomy, University of Science and Technology of China, Hefei 230026, People’s Republic of China\label{instFei}
        \and
        Centro Astron\'omico Hispano en Andaluc\'{i}a, Observatorio de Calar Alto, Sierra de los Filabres, 04550 G\'ergal, Almer\'ia, Spain\label{instCAHA}
        \and
        Departamento de F\'{i}sica de la Tierra y Astrof\'{i}sica 
and IPARCOS-UCM (Instituto de F\'{i}sica de Part\'{i}culas y del Cosmos de la UCM), 
Facultad de Ciencias F\'{i}sicas, Universidad Complutense de Madrid, 28040, Madrid, Spain\label{instUCM}
        \and
        Institut de Ci\`encies de l’Espai (ICE, CSIC), Campus UAB, c/ de Can Magrans s/n, 08193 Bellaterra, Barcelona, Spain\label{instICE}
        \and
        Institut d’Estudis Espacials de Catalunya (IEEC), 08034 Barcelona, Spain\label{instIEEC}
                \and
		Hamburger Sternwarte, Universit\"at Hamburg, Gojenbergsweg 112, 21029 Hamburg, Germany\label{instHS}
                       }

\date{Received 28 September 2023 / Accepted dd Month 2023} 

\abstract
{
To date, the hot Jupiter \wab\ has been the only planet with confirmed orbital decay. 
The late F-type host star has been hypothesized to be surrounded by a large structure of circumstellar material evaporated from the planet.
We obtained two high-resolution spectral transit time series with \carm\
and extensively searched for absorption signals by the atomic species
Na, H, Ca, and He using transmission spectroscopy, thereby covering the \hei\ triplet with high resolution for the first time.
We apply \sys\ for atomic line transmission spectroscopy, introduce the technique of signal protection to improve the results for individual absorption lines, and compare the outcomes to those of established methods.
No transmission signals were detected and the most stringent upper limits as of yet were derived for the individual indicators.
Nonetheless, we found variation in the stellar \ha\ and \hei\ lines, the origin of which remains uncertain but is unlikely to be activity.
To constrain the enigmatic activity state of \wa, we
analyzed \xmm\ X-ray data and found the star to be moderately active at most.
We deduced an upper limit for the X-ray luminosity and the irradiating X-ray and extreme ultraviolet (XUV) flux of \wab.
Based on the XUV flux upper limit and the lack of the \hei\ signal, our hydrodynamic models slightly favor a moderately irradiated
planet with a thermospheric temperature of $\lesssim 12\,000$~K, and a conservative upper limit of
$\lesssim 4\times 10^{12}$~\gs\ on the mass-loss rate.  
Our study does not provide evidence for an extended planetary atmosphere or absorption by circumstellar material close to the planetary orbit. 
}


\maketitle

\section{Introduction}
\label{sec:intro}

The hot Jupiter \wab\ orbits its late F-type host star every $1.09$~d
at a separation of $0.024$~au \citep{Hebb2009}, 
corresponding to a mere $3.1$ stellar radii. The \wa\ planetary system is a member of a larger hierarchical
triplet, including an additional M-dwarf binary at a projected separation of about 1\,arcsec, 
which consists of almost identical stars \citep{Bergfors2013, Bechter2014}.
\wa\ stands out as the only known planetary system that shows transit timing variations, 
widely accepted to originate from
planetary orbital decay caused by tidal interactions \citep[][]{Bailey2019, Yee2020, Turner2021, Wong2022, Bai2022}.
As another promising candidate for orbital decay, WASP--4\,b has been identified only recently \citep[][]{Harre2023}.
The geometric white-light albedo of the planet is very low \citep[$A_{\rm geom} < 0.064$, ][]{Bell2017}.
The thermally dominated phase curve of \wab\ observed by the Transiting Exoplanet Survey Satellite (TESS) satellite shows a marginal eastward offset,
indicative of a superrotating equatorial wind \citep{Wong2022}. The photometry also shows a large contrast
between day and night sides. No significant variability
was found in the TESS photometry \citep[][]{Owens2021, Wong2022}.

\begin{table}[]
    \caption{Relevant system parameters of \wa.
    \label{tab:props}}
    \begin{tabular}{l c l}
    \hline\hline
    \noalign{\smallskip}
    Parameter & Value & Source$^a$ \\
    \noalign{\smallskip}
    \hline
    \noalign{\smallskip}
    \rs\ [\rsun] & $1.66\pm 0.05$ & T21$^b$ \\
    \rprs & $0.1166445 \pm 0.0000097$ & T21 \\
    \rp\ [\rj] & $1.884\pm 0.057$ & T21 \\
    $a$ [au] & $0.02399 \pm 0.00072$ & T21 \\
    \mp\ [\mj] & $1.46\pm 0.27$ & T21 \\
    $\rho_{\rm p}$ [g\,cm$^{-3}$] & $0.271\pm 0.056$ & T21 \\
    $\log{g_p}$ [cgs] & $3.03\pm 0.04$ & TW$^c$ \\
    $i_{\rm orb}$ [deg] & $84.955\pm 0.037$ & T21 \\
        $d_{\star}$ [pc] & $413.0\pm 3.3$ & DR3 \\
    \tc{eff} [K] & $6250\pm 100$ & F10 \\
    \tc{eq} [K] & $2520\pm 40$ & TW$^d$ \\
    $v\sin{i}$ [\kms] & $1.6_{-0.4}^{+0.8}$ & A12 \\
    $\gamma$ [\kms] & $19.46\pm 0.02$ & TW \\
    $K_{\star}$ [\ms] & $226\pm 4$ & H09 \\
    $\log{R_{\rm HK}'}$ & $-5.5$  & K10 \\
    \noalign{\smallskip}
    \hline
    \noalign{\smallskip}
    \multicolumn{3}{c}{Ephemerides (constant period)}\\
    $T_0$ [BJD$_{\rm TDB}$] & $2456305.455519\pm 0.000026$ & T21 \\
    $P_{\rm orb}$ [d] & $1.091419426\pm 0.000000022$ & T21 \\
    \noalign{\smallskip}
    \multicolumn{3}{c}{Ephemerides (orbital decay)}\\
    $T_0$ [BJD$_{\rm TDB}$] & $2456305.455795\pm 0.000038$ & T21 \\
    $P_{\rm orb}$ [d] & $1.091420090\pm 0.000000041$ & T21 \\
    $dP_{\rm orb}$\,d$E$$^{-1}$ [d/orbit] & $(-9.45\pm 0.47)\times 10^{-10}$ & T21 \\
    \noalign{\smallskip}
    \hline
    \end{tabular}
\tablefoot{$^a$ T21: \citet{Turner2021}, F10: \citet{Fossati2010b}, DR3:
Gaia Data Release 3 \citep{Gaia2016A, Gaia2023}, 
A12: \citet{Albrecht2012}, 
H09: \citet{Hebb2009},
K10: \citet{Knutson2010}, 
TW: This work.
$^b$ Value adopted by T21 (derived from their numbers). The value is consistent with that of $1.59$~\rsun\ reported by \citet{Stassun2017} and the radii implicit in the model masses and densities given by \citet{Bailey2019}, and the radius of $1.57\pm 0.07$~\rsun\ reported by \citet{Hebb2009}.
$^c$ Calculated using Eq.~4 by \citet{Southworth2007}.
$^d$ Assuming zero Bond albedo (Eq.~\ref{eq:teq}).}
\end{table}

The activity state of \wa\ remains enigmatic. On the one hand,
the star shows an unexpectedly low spectroscopic $v\sin{i}$ value \citep[Table~\ref{tab:props} --][]{Fukuda1982, Albrecht2012}
and unusually dark cores in the chromospherically sensitive \ion{Mg}{II} and \ion{Ca}{II}
lines \citep{Haswell2012, Fossati2013}, which may point to an extremely low
activity level. On the other hand, \citet{Haswell2012} also report a likely flaring event, and
the sky-projections of the planetary orbit normal and the stellar spin axis
are misaligned by $59_{-20}^{+15}$\,deg \citep{Albrecht2012}, so that a projection
effect is a plausible explanation for the low $v\sin{i}$ value.
The exceptionally low flux in the cores of the chromospheric lines of \wa\ 
may be explained by absorption in a very extended structure of circumstellar material originating from the
planet \citep{Haswell2012}, which may be a phenomenon also occurring in other planetary systems
with close-in planets \citep{Fossati2013}. 
The one-dimensional atmospheric simulations by \citet{Salz2016a, Salz2016b}
yield a mass-loss rate of $4\times 10^{11}$~g\,s$^{-1}$ for \wab, which is
about an order of magnitude higher than the value adopted by \citet{Haswell2012} and in that sense
consistent with the formation of an extended planetary-fed structure.
The three-dimensional gas-dynamical modeling by \citet{Bisikalo2013, Bisikalo2015} indicates that
the material mainly escapes in the form of two streams across the first and second Lagrange points.
Finally, the 3D hydrodynamic simulations by \citet{Debrecht2018} also show the formation of a two-streamed
outflow morphology along with the development of an extended circumstellar torus, which reaches sufficient
densities to affect strong spectral lines such as the \ha\ line and potentially the \hei\ triplet
on the timescale of only a few years.   
Altogether, it appears that \wa\ is unlikely
to be extremely inactive.

Its unusual properties have made \wab\ a coveted target with studies for emission and transmission spectroscopy.
While the emission of \wab\ has been analyzed in some detail, many aspects of its (dayside) atmospheric
structure remain under discussion. In particular, it is debated
if and to what extent the carbon-to-oxygen ratio is elevated and if a thermal
inversion layer exists as one expects in planets such as \wab\
\citep{Madhusudhan2011, Stevenson2014b, Line2014, Oreshenko2017, Himes2022}.

Likewise, numerous studies of the transmission spectrum exist.  
\citet{Fossati2010}, \citet{Haswell2012}, and \citet{Nichols2015}
present near-ultraviolet transmission spectroscopy of a total of six primary transits of
\wab\ observed with the Cosmic Origins Spectrograph (COS) onboard the Hubble Space
Telescope (HST). These data reveal excess in-transit absorption \citep{Nichols2015} with
increased transit depths observed in resonance lines of
various metals \citep{Fossati2010}. The near-ultraviolet transit observations show variability and are
indicative of an extended atmospheric structure overfilling the planetary Roche lobe 
\citep{Haswell2012}. 

\citet{Sing2013} study the transmission spectrum of \wa\ in the 2900--10300\,{\AA} range at low spectral resolution.
They find their transmission spectrum to show a Rayleigh scattering slope while being otherwise
devoid of any strong molecular or atomic absorption features. Specifically, the authors
report non-detections of Na, K, \ha, and H$\beta$ using $30$~\AA-wide
resolution elements, and conclude that, if present at all, such features
must be confined to a narrow core region. Indeed,
\citet{Burton2015} and \citet{Bai2022} find a tentative detection of excess in-transit absorption in the \nad\ lines in the
spectra of \wa, and
\citet{Jensen2018} describe pronounced excess transit absorption in the \nad\ and \ha\ lines of \wa\ observed
at moderately high spectral resolution ($R\approx 15\,000$) but with noncontinuous temporal sampling, possibly
extending into the pre- and post-transit phases. \citet{Kreidberg2018} analyze data obtained by the Wide Field
Camera 3 onboard the Hubble Space Telescope and report the non-detection
of \hei\ absorption, however, at low spectral resolution.
Analyzing the transmission spectrum up to $5$~\mbox{$\mu$m},
\citet{Stevenson2014} report features likely attributable to molecular absorption and
\citet{Kreidberg2015} detect water in the transmission spectrum of \wab.
The atmosphere of \wab\ is thought to have a cloud deck, which attenuates spectral
features from the underlying atmosphere \citep{Wakeford2017}, which can explain the apparent lack
of strong atmospheric signals.

\section{Observations}

We obtained two spectral transit time series of \wa\ with CARMENES\footnote{Calar Alto high-Resolution search for M dwarfs with Exoearths with Near-infrared and optical Échelle Spectrographs} \citep{Quirrenbach2018}.
The CARMENES spectrograph has two channels, namely VIS and NIR, that cover the wavelength ranges $0.52$--$0.96$~$\mu$m\ and $0.96$--$1.71$~$\mu$m, respectively. 
The spectral resolutions are $94\,600$ in the VIS channel and $80\,400$ in the NIR channel.

Our two observing runs commenced on 18 December 2019 20:04
and 17 November 2020 22:00 (UTC), which we refer to as night~1 and night~2 in the following. 
A total of 22 and 21 spectra with individual exposure times of 15~minutes where obtained during our runs.
The data were reduced using the \texttt{caracal} pipeline
\citep[CARMENES Reduction And CALibration --][]{Zechmeister2014, Caballero2016}
and a telluric correction was subsequently performed using the \texttt{molecfit} package
\citep{Smette2015, Kausch2015}.
Figure~\ref{fig:airmass} shows the evolution of airmass and S/N per pixel in the
order covering the H$\alpha$ line along with the transit contact points.

\subsection{Systemic velocity}
\label{sec:sysrv}

To determine the systemic radial velocity, $\gamma$, of the system, we first obtained
master spectra for the individual spectral orders of the CARMENES VIS channel. To that end,
we carried out the barycentric correction for the telluric-corrected spectra and obtained the
pixel-wise median spectrum. The thus obtained    
master order spectra were then compared to a synthetic PHOENIX spectrum computed for $T_{\rm eff} = 6200$~K and $\log g=4.5$ \citep{Husser2013}. The comparison was done by a fit in which
we considered a free normalization by a quadratic polynomial, the coefficients of which were treated as nuisance parameters,
and a free parameter for the systemic radial velocity (RV).
We then calculated the median
value and the median deviation about the median as a robust estimator of the standard deviation
\citep[][]{Hampel1974, Rousseeuw1993, Czesla2018}, which we subsequently divided by the square root
of the number of orders to get the error of the mean.
In Table~\ref{tab:props}, we list the thus obtained best-fit value for the systemic velocity.
Our result of $19.46 \pm 0.02$~\kms is well consistent yet more precise than the value of $20.6\pm 1.4$~\kms\ reported by \gaia\ DR3.

\subsection{Roche geometry}
\label{sec:Roche}

\begin{figure}
        \includegraphics[trim=50 0 50 0,clip, width=0.49\textwidth]{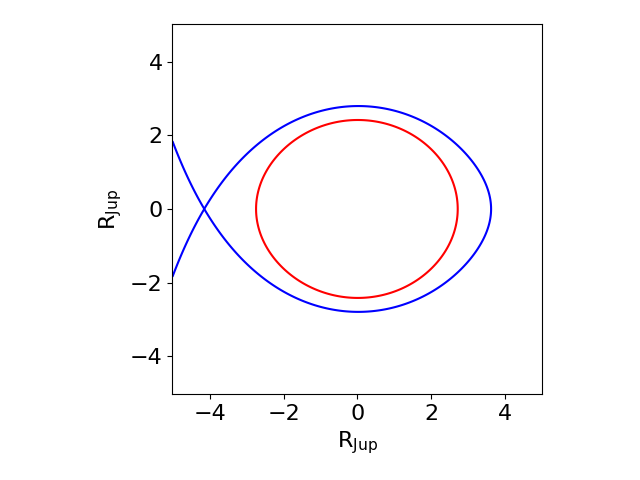}
    \caption{Pole-on view of the Roche geometry of \wab. The host star is to the left.
    The equipotential surfaces of the
    Roche lobe (blue) and planetary surface (red) are indicated.    \label{fig:roche}}
\end{figure}

\wab\ is so close to its host star that tidal forces become significant. The combined gravitational and centrifugal
forces on test masses in the rotating frame of two orbiting bodies is described by the Roche potential
\citep[e.g.,][and App.~\ref{sec:rochePot}]{Hilditch2001}.     
In Fig.~\ref{fig:roche}, we show a pole-on view (i.e., the line-of-sight is along the normal of
the orbital plane) of \wab's geometry, indicating the equipotential surfaces defining the Roche lobe
and the planetary body, which both assume typical drop-like shapes.
While the deviation from the spherical shape is strongest in the substellar direction, also the polar and side
radii of the drop differ slightly \citep[e.g.,][]{Haswell2012}. As photometric transit measurements are primarily sensitive to
the fractional area covered by the occulting planet, we interpret the
planetary radius, \rp, given in Table~\ref{tab:props} as the radius of a circular disk with the same area
as the projection of the drop-like planetary body during conjunction. Modeling the latter as an ellipse, we obtain 
the potential level defining the planetary surface by the condition of equal area \citep[see, e.g.,][]{Delrez2016}
\begin{equation}
    \mbox{\rp}^2 = R_{\rm p, pole} \times R_{\rm p, side} .
\end{equation}
This criterion yields the following values
for the radii in the substellar, polar, and side directions: 
$R_{\rm p, subsolar} = 2.18$~\rj, $R_{\rm p, pole} = 1.86$~\rj, and $R_{\rm p, side} = 1.91$~\rj.
The corresponding radii
of the Roche lobe read $R_{\rm Rl, subsolar} = 3.29$~\rj, $R_{\rm Rl, pole} = 2.12$~\rj,
and $R_{\rm Rl, side} = 2.21$~\rj.
The planetary body fills $59$\,\% of its Roche lobe volume. 

\subsection{Transit timing}
\label{sec:tt}

Nights~1 and 2 covered the transits with epochs no. 2319 and 2626 with respect to the
ephemerides given by \citet{Turner2021} (Table~\ref{tab:props}).
While the change in orbital period predicted by the orbital decay model is only about $-0.2$~s per epoch, the difference in the
transit times obtained using the constant period and orbital decay ephemerides
accumulates to between one and two minutes at these epochs.
A two minute timing error amounts to a RV shift of $1.9$~\kms\ in the planetary rest frame during the
center of the transit and, thus,
transmission spectrum, if not accounted for.  
Throughout the paper, we adopt the orbital decay ephemerides published by \citet{Turner2021}.
The more recent quadratic ephemerides provided by \citet{Wong2022} and \citet{Bai2022} yield consistent
transit times, which differ by only a few seconds for our events.

Orbital motion causes an acceleration of $15.9$\,m\,s$^{-2}$ of the planet along the line-of-sight direction during the central transit phase.
For individual exposure times of $900$~s, this results in a change of $14.3$~\kms\ 
in the orbital planetary RV during every single exposure.
Using the formulation given by \citet{Czesla2022}, we treated the broadening effect of this
phase smearing on the transmission spectrum by adopting an effective instrumental resolution of $30\,000$, which is
appropriate
for both channels of CARMENES, because the smearing dominates the line broadening. 

\begin{figure}
		\includegraphics[width=0.49\textwidth]{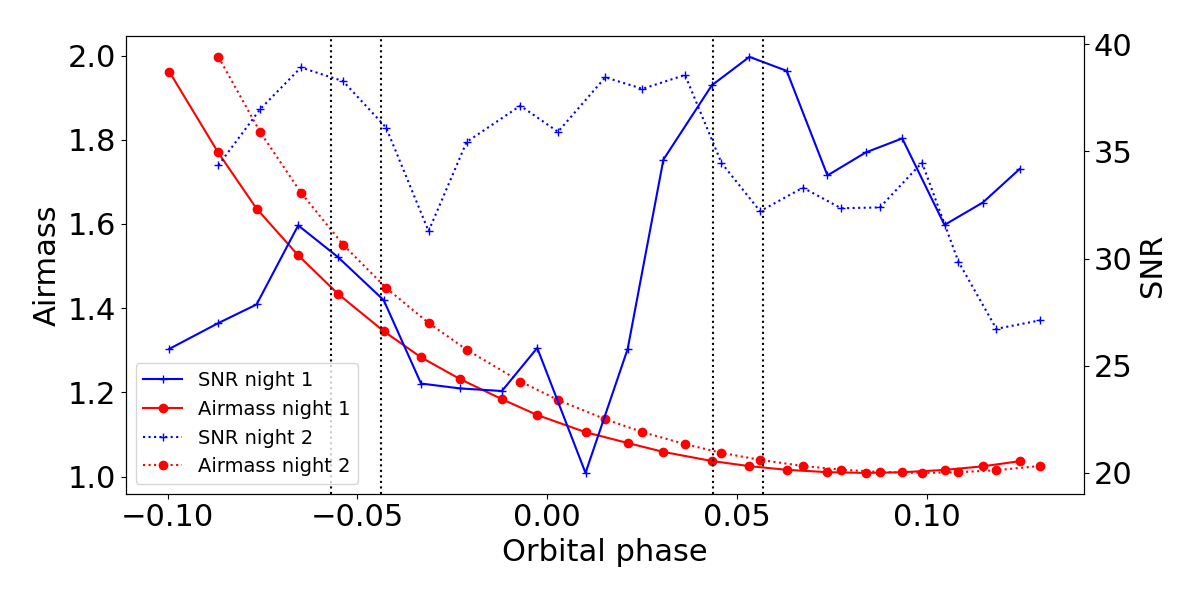}
	\caption{Evolution of airmass and S/N in night~1 (solid) and night~2 (dotted).
	Dotted vertical lines indicate the four contact points of the transit.
	\label{fig:airmass}}
\end{figure}

\subsection{X-ray observations with \xmm}
\label{sec:xmm}

We observed \wa\ with \xmm\ on 14 September 2019 (PI: Sanz-Forcada).
Our 9~ks exposure with the combined EPIC detectors only gave a poor S/N=1.3
detection (Sanz-Forcada et al. in prep). The large stellar distance
implies a canonical interstellar medium \ion{H}{i} absorption column of $\approx 2\times
10^{21}$~cm$^{-2}$. We performed a spectral fit to the source, forcing a
``solar'' coronal temperature of $\log~T\mbox{(K)}=6.3$, which yields an upper
limit of $6\times 10^{28}$~erg\,s$^{-1}$ in the X-ray (5--100~\AA) band.
The relatively low ratio of X-ray to bolometric luminosity, $\log L_{\rm X}/L_{\rm bol}
< -5.4$, indicates that the star is of moderate activity at most,
consistent with observations of the \ion{Ca}{II} lines. The resulting upper limit
of $4.5\times 10^{29}$~erg\,s$^{-1}$ on the extreme
ultraviolet (EUV, 100--920~\AA) stellar flux was calculated with a coronal plasma model extrapolated
to the temperatures of the transition region as outlined by \citet{Sanz2011}. While X-ray--EUV scaling relations
such as those by \citet{Chadney2015}, \citet{King2018}, \citet{Johnstone2021}, or, in fact,
\citet{Sanz2011} themselves may provide slightly different estimates, we emphasize that the main source
of uncertainty is the poor X-ray photon statistics.

\section{Analysis and results}
\label{sec:analysis}
In the following, we study the use of \sys\ in atomic line spectroscopy (Sect.~\ref{sec:sysremTS}) and apply the technique
to \wa\ (Sect.~\ref{sec:ts}). We then analyze spectral changes in the stellar reference frame (Sect.~\ref{sec:internight})
and examine interstellar absorption (Sect.~\ref{sec:na_ism}).

\subsection{Atomic line transmission spectroscopy with \sys}
\label{sec:sysremTS}

The observed spectra contain the stellar and planetary signals
as well as a number of confounding components. 
The most important of these in ground-based transmission spectroscopy are the presence of
potentially variable telluric absorption and emission lines as well as interstellar absorption lines.
Additionally, stellar line profile variations such as those caused by activity and rotation may exist.

\subsubsection{Representation of time series measurements}

When a spectral time series is observed with a telescope, measurements of the specific flux from some solid angle in the sky
are obtained in $N$ time intervals, $[t_i, t_i+\Delta t_i]$,
and $M$ wavelength bins, $[\lambda_j, \lambda_j+\Delta \lambda_j]$.
For the purpose of the following discussion, we assume that the specific flux, $\mathcal{F}_{\rm T}(\lambda, t)$,
collected from the direction of the transiting planetary system can be modeled by the following expression 
\begin{align}
    \mathcal{F}_{\rm T}(\lambda, t) = 
    \mathcal{F}_{\star}(\lambda, t)  e^{-\tau_{\rm ISM}(\lambda, t) -\tau_{\rm TA}(\lambda, t)
    -\tau_{\rm p}(\lambda, t)} + \mathcal{F}_{\rm TE}(\lambda, t) .
    \label{eq:specrep}
\end{align}
Here $\mathcal{F}_{\star}(\lambda, t)$ is the disk-integrated stellar flux scaled by distance, $\tau_{\rm ISM}(\lambda, t)$
is the optical depth of the interstellar medium, $\tau_{\rm TA}(\lambda, t)$ is the optical depth of
telluric absorption, $\tau_{\rm p}(\lambda, t)$ is the optical depth
of the planetary atmosphere in front of the stellar disk, and $\mathcal{F}_{\rm TE}(\lambda, t)$ represents the flux of telluric emission
features observed in the direction of the target.
The optical depth of the planetary atmosphere $\tau_{\rm p}(\lambda, t)$ is often only larger than zero during the optical transit, but
examples of prolonged or early starting atmospheric transits are known \citep[e.g.,][]{Cauley2015, Nortmann2018, Czesla2022, Zhang2023}.
Equation~\ref{eq:specrep} remains an approximation at least because the planetary opaque disk blocks out some light and
the absorption by the planetary atmosphere is a local phenomenon, as it
does not act upon the disk-integrated spectrum \citep[e.g.,][]{Dethier2023}.
If spaceborne observatories such as the Hubble
Space Telescope are used, the telluric absorption term in Eq.~\ref{eq:specrep} vanishes (i.e., $\tau_{\rm TA}(\lambda, t) = 0$). Yet, emission terms
such as those attributable to geocoronal emission lines may remain significant.
We assume that the telescope system and data reduction can be represented by the function $\mathcal{I}$
\begin{align}
	\mathcal{I} : \mathcal{F}_{\rm T}(\lambda, t) \mapsto \{F_{i,j} | i=1\ldots N, j=1\ldots M \} \; ,
\end{align}
where the $F_{i,j}$ are normally distributed random variables with expected values given by
\begin{align}
	\mu_{i,j} = \frac{1}{\Delta\lambda_j\, \Delta t_i} \int_{\lambda_j}^{\lambda_j+\Delta \lambda_j} \int_{t_i}^{t_i+\Delta t_i}
	\mathcal{R} \left( \mathcal{F}_{\rm T}(\lambda, t)*\mathcal{P}(\lambda) \right)
	\, d\lambda \, dt \;
\end{align}
and standard deviations denoted by $\sigma_{i,j}$. 
The $*$ symbol implies convolution with the instrumental profile, $\mathcal{P}(\lambda)$,
and the function $\mathcal{R}$ summarizes the spectral response of the instrumental system and
data reduction process. 

Actual measurements are considered to be realizations, $f_{i,j}$, of the random variables $F_{i,j}$.
A time series observation may be represented by $N\times M$ matrices
\begin{align}
	\vec{F} = \begin{pmatrix} f_{1,1} & \ldots & f_{1,M} \\
	\ldots &  & \ldots \\
	f_{N,1} & \ldots & f_{N,M} \\
	\end{pmatrix} \;\;
	\mbox{and}
	\;\;
	\vec{\Sigma} = \begin{pmatrix} \sigma_{1,1} & \ldots & \sigma_{1,M} \\
	\vdots &  & \vdots \\
	\sigma_{N,1} & \ldots & \sigma_{N,M} \\
	\end{pmatrix} \; ,
	\label{eq:matrixrep}
\end{align}
where the rows of $\vec{F}$ are the observed spectra and $\vec{\Sigma}$ holds the uncertainties. 

While the relative RV shifts of the spectral components included in Eq.~\ref{eq:specrep} are defined by
their physical origin, their absolute shifts and to some extent their time evolution
are determined by the reference frame with respect to which
the spectra are given. 
The most relevant reference frames for our purpose are the observer's reference frame (OF), the barycentric
reference frame (BF), the stellar reference frame (SF), and the planetary reference frame (PF).
Observations are naturally obtained in the OF, although the data reduction pipeline of the
spectrograph may already apply a transformation during the extraction of the one-dimensional spectra from the raw frames.
The telluric absorption and emission lines remain stationary in the OF of a ground-based observatory
\citep[at least to within about $20$~\ms, ][]{Caccin1985}, but their strengths are usually time dependent.

Transformations of the matrices $\vec{F}$ and
$\vec{\Sigma}$ between reference frames can be obtained by applying appropriate Doppler shifts
to the individual rows. For example, we denote by
\begin{equation}
    \vec{F}_{\rm BF} = \mathcal{D}\left(\vec{F}_{\rm OF}; \vec{v}_{\rm BC} \right),
\end{equation}
a transformation of the matrix $\vec{F}_{\rm OF}$ from the OF into the BF by applying the
appropriate barycentric correction, $v_{i, \rm BC}$, to the $i^{\rm th}$ row of $\vec{F}_{\rm OF}$;
the corrections for all rows are
given by the vector $\vec{v}_{\rm BC} = (v_{1, \rm BC} \ldots v_{N, \rm BC})$.

\subsubsection{The \sys\ algorithm}
\label{sec:sysrem}

The \sys\ algorithm was described by \citet{Tamuz2005}. It can be applied to matrices of the form
\begin{align}
    \vec{R} = \begin{pmatrix} r_{1,1} & \ldots & r_{1,M} \\
    \ldots &  & \ldots \\
    r_{N,1} & \ldots & r_{N,M} \\
    \end{pmatrix}
    \;\;
    \mbox{and}
    \;\;
    \vec{\Sigma}_{\rm R} = \begin{pmatrix} \sigma_{{\rm R}, 1,1} & \ldots & \sigma_{{\rm R}, 1,M} \\
    \vdots &  & \vdots \\
    \sigma_{{\rm R}, N,1} & \ldots & \sigma_{{\rm R}, N,M} \\
    \end{pmatrix}
    \label{eq:sysrem_matrices}
\end{align}
where $\vec{R}$ contains the input data and $\vec{\Sigma}_{\rm R}$
the standard deviations, $\sigma_{{\rm R},i,j}$, of the associated uncertainties.
\sys\ defines a mapping, $\mathcal{S}$, from $\vec{R}$ and $\vec{\Sigma}_{\rm R}$ to a model, $\vec{M}$,
\begin{align}
    \mathcal{S} : (\vec{R}, \vec{\Sigma}_{\rm R}) \mapsto \vec{M} \; ,
\end{align}
which approximates the input data.
The model is the outer product of two (column)
vectors $\vec{a} = (a_1, \ldots, a_N)$ and $\vec{c} = (c_1, \ldots, c_M)$
\begin{align}
    \vec{M} =  a_i \, c_j = \vec{a} \vec{c}^{\rm T} \; ,
    \label{eq:sysremmodel}
\end{align}
the elements of which are defined by the condition that the quantity
\begin{align}
    \sum_{i,j} \frac{\left(r_{i,j} - a_i \, c_j \right)^2 }{\sigma_{{\rm R},i,j}^2} \; ,
\end{align}
is minimized. The input data matrix is then updated by subtracting the model
\begin{align}
    \vec{R}^{(k+1)} = \vec{R}^{(k)} - \vec{M}^{(k)} = \vec{R}^{(k)} - \vec{a}^{(k)} \vec{c}^{(k)\rm T} \; .
    \label{eq:sysremcorr}
\end{align}
As the \sys\ algorithm is typically applied iteratively, we follow \citet{Tamuz2005} and use
the superscript $(k)$ to denote the number of iterations and understand that $\vec{R}^{(0)} = \vec{R}$.
From Eq.~\ref{eq:sysremcorr}, it follows that after $k$ iterations
\begin{equation}
    \vec{R}^{(k)} = \vec{R}^{(0)} - \sum_{l=0}^{k-1} \vec{M}^{(l)} = \vec{R}^{(0)} - \mathbfcal{M}^{(k)} \; .
\end{equation}
We call $\mathbfcal{M}^{(k)}$ the cumulative model and refer to the rows of $\vec{R}^{(k)}$ as residual spectra.

In the analysis of spectral time series with \sys, 
one possibility is to use the individual spectra (or sections thereof) in the OF as rows of the data matrix,
that is, $\vec{R}^{(0)} = \vec{F}_{\rm OF}$ and $\vec{\Sigma}_{\rm R} = \vec{\Sigma}_{\rm OF}$.
Intuitively, the first \sys\ model ($\vec{M}^{(0)}$, Eq.~\ref{eq:sysremmodel}) then consists of
a kind of average spectrum, $\vec{c}^{(0)}$, with 
individual scaling factors, $a_i$, to best match the observations. Such a model
mainly reproduces components with stationary wavelengths and variable strength, but does not efficiently 
reproduce components, shifting fast through the spectra.
Therefore, the suitability of the reference frames for the analysis
is largely determined by the ensuing wavelength shifts of the individual components in Eq.~\ref{eq:specrep}.

For close-in planets, orbital motion can produce strong RV variations of the
planetary component with regard to the stellar and telluric spectral components,
which are crucial to distinguish planetary signals from other effects.
During our two individual observing runs targeted at \wa,
the barycentric correction changes by $\lessapprox 0.5$~\kms\ and the stellar reflex motion produces a Doppler
shift of \mbox{$\lessapprox 0.3$~\kms}. In contrast, the RV induced by the planetary orbital motion changes 
by about $160$~\kms.

\subsubsection{Deriving the observable transmission spectrum}
\label{sec:sysrem_estimator}

We call the quantity
\begin{align}
	f_{\rm t} (\lambda, t) = \left(e^{-\tau_p(\lambda, t)} - 1\right) * \mathcal{P}(\lambda)
\end{align}
the observable transmission spectrum, which is the planetary component in Eq.~\ref{eq:specrep} with
unity subtracted, broadened by the instrumental profile.
Our problem is to define an estimator
for the observable transmission spectrum.
This estimator should have favorable properties such as being
unbiased.

Based on $\vec{R}^{(k)}$ and $\vec{\Sigma}_{\rm R}$, expressed in any reference frame,
average residual spectra, $\vec{b} = (b_1, \ldots, b_M)$, can be constructed by
defining a function $\mathcal{B}$ such that 
\begin{align}
    \mathcal{B}: \vec{R}, \vec{\Sigma}, \mathcal{J} \mapsto \vec{b} \;\;\;
    \mbox{with} \;\;\;
    b_j = \frac{\sum_{i \in \mathcal{J}} r_{i,j} \sigma_{i,j}^2}{\sum_{i \in \mathcal{J}} \sigma_{i,j}^2} ,
    \label{eq:ars}
\end{align}
where $\mathcal{J}$ is the set of indices of the rows considered in the averaging.
Thus, for example,
$\mathcal{B}(\vec{F}_{\rm OF}, \vec{\Sigma}_{\rm OF}, \mathcal{J})$
is the error-weighted, bin-wise averaged observed spectrum over the indices $\mathcal{J}$ in the OF.

Assume, we start the \sys\ iteration with the observed spectral time series expressed in a hypothetical
reference frame, XF, in which all but the planetary signal can be assumed to be stationary 
($\vec{R} = \vec{F}_{\rm XF}$ and $\vec{\Sigma}_{\rm R} = \vec{\Sigma}_{\rm XF}$).
As an idealization, we may then assume that
after a sufficient number, $k_{\rm s}$, of \sys\ iterations,
all spectral components in Eq.~\ref{eq:specrep} but the fast-shifting planetary signal
are reproduced by the model so that the expected value of the cumulative model becomes
\begin{equation}
	E\left[ \mathbfcal{M}^{(k_s)}\right] = E\left[ \mathcal{I}(
	\mathcal{F}_{\star}(\lambda, t) e^{-\tau_{\rm ISM}(\lambda, t) -\tau_{\rm TA}(\lambda, t)} +
	\mathcal{F}_{\rm TE}(\lambda, t)) \right] \; .
\end{equation}
Subtracting the cumulative model from the observations gives the matrix $\vec{R}^{(k_s)}$ with expected value
\begin{align}
	E\left[\vec{R}^{(k_s)}\right] = E\left[\vec{F} - \mathbfcal{M}^{(k_s)} \right] \, = \nonumber \\
	E\left[ \mathcal{I}(\mathcal{F}_{\star}(\lambda, t) e^{-\tau_{\rm ISM}(\lambda, t) -\tau_{\rm TA}(\lambda, t)} 
	\left(e^{-\tau_{\rm p}(\lambda,t)} - 1 \right) 
	 \right] \; .
\end{align}
If $\mathcal{J}_{\rm IN}$ denotes the set of in-transit indices,
\begin{equation}
\mathcal{B}(\vec{R}_{\rm PF}^{(k_s)}, \vec{\Sigma}_{\rm PF, R}, \mathcal{J}_{\rm IN}) 
\end{equation}
is the average in-transit residual spectrum in the PF, which is an estimator of the observable transmission spectrum averaged over the in-transit
period. It is, however, not generally
an unbiased estimator of $f_t(\lambda, t)$. Only if the factor $\mathcal{F}_{\star}(\lambda, t) e^{-\tau_{\rm ISM}(\lambda, t) -\tau_{\rm TA}(\lambda, t)}$
is unity, implying a flat stellar spectrum and a continuum normalization, the estimator is unbiased.
The basic problem here is that the basic \sys\ operation is a subtraction, while the
planetary atmospheric absorption component is a multiplicative factor acting upon the underlying continuum. 

Assuming that no telluric emission is present ($\mathcal{F}_{\rm TE}(\lambda, t)) = 0$), a more general, ideally unbiased estimator can be constructed
if we divide by the cumulative model
\begin{align}	
	\mathcal{T} : \vec{R}_{\rm XF}, \mathbfcal{M}^{(k_s)}_{\rm XF}, \vec{\Sigma}_{\rm XF, R}, \vec{v}_{\rm XF}, \mathcal{J}_{\rm IN} \mapsto \nonumber \\
	B\left( \mathcal{D}\left( \frac{\vec{R}_{\rm XF}}{\mathbfcal{M}^{(k_s)}_{\rm XF}} ,\vec{v}_{\rm XF}\right),
	\mathcal{D}\left(\frac{\vec{\Sigma}_{\rm XF, R}}{\mathbfcal{M}^{(k_s)}_{\rm XF}}, \vec{v}_{\rm XF}\right) ,
	\mathcal{J}_{\rm IN}\right),
	\label{eq:estimator_T}
\end{align}
where the matrix division is to be understood element-wise and $\vec{v}_{\rm XF}$ denotes the velocities required to transform the matrices into
the planetary frame. In words, evaluating the estimator $\mathcal{T}$ requires that we (i) apply \sys\ for $k_s$ iterations,
(ii) divide the input matrices element-wise by the cumulative model, (iii) transform the matrices into the planetary frame, and (iv) compute
the average residual spectrum including the desired set of spectra.
Ideally,
the expected value of the estimate, $\tau_j$, for the $j^{\rm th}$ spectral bin is given by
\begin{equation}
	E\left[\tau_{j} \right] = \frac{1}{\Delta\lambda_j} \int_{\lambda_j}^{\lambda_j+ \Delta \lambda_i}
	\left( e^{-\tau_p(\lambda)} - 1\right) * \mathcal{P}(\lambda) d\lambda ,
	\label{eq:expT}
\end{equation}
making $\mathcal{T}$ a good estimator of the observable transmission spectrum. 
This implicitly assumes that convolution
is distributive over multiplication so that we may write
\begin{align}
\mathcal{P}(\lambda) * \left(\mathcal{F}_{\star}(\lambda, t) e^{-\tau_{\rm ISM}(\lambda, t) -\tau_{\rm TA}(\lambda, t)} \left(e^{-\tau_{\rm p}(\lambda,t)} - 1 \right)\right) = \nonumber \\  
\mathcal{P}(\lambda) * \left(\mathcal{F}_{\star}(\lambda, t) e^{-\tau_{\rm ISM}(\lambda, t) -\tau_{\rm TA}(\lambda, t)} \right) \times \mathcal{P}(\lambda) * \left(e^{-\tau_{\rm p}(\lambda,t)} - 1 \right),
\end{align}
which is not the case, but a reasonable approximation in our problem (see App.~\ref{sec:convolution_distributive}). 
The methodology represented by
Eq.~\ref{eq:estimator_T} is essentially that used by \citet{Gibson2020}, who correctly asserted that it ``preserves the relative
depths of the planet's transmission spectrum'' without giving, however, a more detailed explanation of 
this statement. The technique has since been regularly applied in the literature
\citep[e.g.,][]{Merritt2020, Yan2020, Cont2022, Gibson2022}.

\subsubsection{Signal protection via error adjustment}
\label{sec:signal_protection}

The assertion of Eq.~\ref{eq:expT} depends on more idealizations than the distributivity of convolution.
In particular, the \sys\ model may also be expected to pick up on the fast shifting
planetary signal although not as efficiently as on more stationary components \citep[e.g.,][]{Brogi2019, Gibson2022}.
A simulation with a simplified though realistic setting readily demonstrates how the \sys\ model starts to
reproduce the shifting absorption signal as the number of iterations is increased. 
For the simulation, we adopt the temporal and spectral sampling of the \ha\ line region
obtained during our night~1 CARMENES observations of \wa. We then simulate synthetic spectra by using
a mock stellar \ha\ line, modeled as a Gaussian centered at $6565.058$~\AA\ with a full width at half maximum (FWHM) of 2~\AA\ and a depth of 0.9,
and subsequently inject a planetary absorption signal at the nominal radial velocity of the planetary orbital
motion. The latter signal is modeled by a Gaussian with a constant equivalent width (EW) of 20~m\AA\ during the transit and a FWHM of $0.15$~\AA.
We include in our simulation the shifts caused by the stellar
orbital motion and the barycentric motion of \wa\ in the OF and add Gaussian noise to simulate a S/N of 100
per spectral bin.
The thus obtained mock spectral time series are represented by the matrices $\vec{F}_{\rm M, OF}$
and $\vec{\Sigma}_{\rm M, OF}$, which we use as the input for the \sys\ algorithm. After applying
\sys\ for $k$ times, we obtain the average in-transit transmission spectrum using Eq.~\ref{eq:estimator_T}
and determine the wavelength, width, and depth of the resulting signal by fitting a Gaussian. Additionally, 
a free continuum offset is included in the fit, which is treated as a nuisance.
By repeating this experiment 1000 times, we 
find the expected values and standard deviations for the parameters, which we list in 
Table~\ref{tab:sysrem_mock}.
The upper section of Table~\ref{tab:sysrem_mock} clearly shows that the EW of the remaining signal in the
reconstructed transmission spectrum decreases as the number of \sys\ iterations increases. It should also be
noted, however, that the width and location of the signal, as obtained by the Gaussian fit, remain intact.

\begin{figure}[]
		\includegraphics[width=0.49\textwidth]{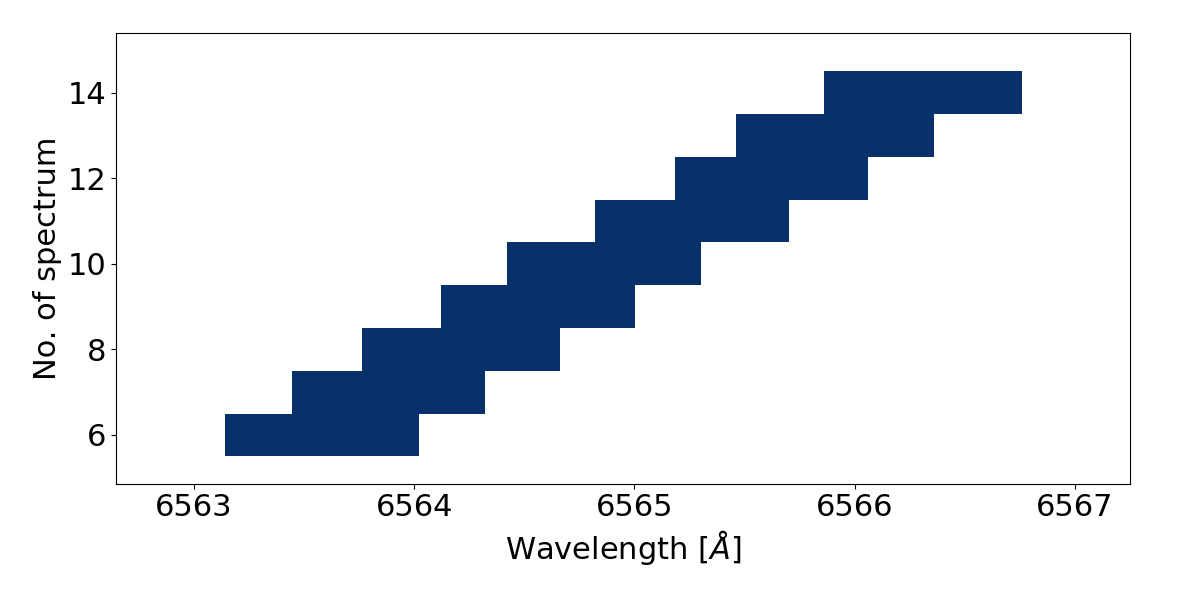}
	\caption{Section of mock uncertainty matrix, $\vec{\Sigma}_{\rm M, OF, SP}$, showing the effect of SP.
	Dark regions indicate increased uncertainties.
	\label{fig:mock_emap}}
\end{figure}

If the planetary orbital velocity is known
with sufficient accuracy, also the likely wavelength of a potential planetary atmospheric absorption signal
can be derived. In this case, the signal can be ``protected from the model'' by modifying the
uncertainties used by \sys\ in signal protection (SP) regions.
To prevent the \sys\ model from picking up on the signal, we strongly increase the
uncertainties of the spectral bins in a region centered on the nominal location of the planetary signal
for all in-transit spectra, making them infinite for practical purposes. We here use a region with a total width of
$\pm 6$~resolution elements. The relevant section of the resulting modified uncertainty matrix,  $\vec{\Sigma}_{\rm M, OF, SP}$, is
shown in Fig.~\ref{fig:mock_emap}, where we used an arbitrary, large value of $10^3$ (i.e., a S/N of $10^{-3}$ per spectral bin)
for the uncertainty in the SP regions.
We emphasize, however, that to evaluate $\mathcal{T}$ (Eq.~\ref{eq:estimator_T}) the original
uncertainties should be used irrespectively. Increased uncertainties in Eq.~\ref{eq:estimator_T} may be used
to mask specific spectral regions, which is regularly done, but not our goal here.  
The effect of SP on our mock problem is demonstrated in the second half of Table~\ref{tab:sysrem_mock}, which
shows that the characteristics of the input signal can be recovered for all tested numbers of \sys\ iterations.

The details of the optimal SP procedure depend on the observation. If, for example, a
prolonged transit is expected, it may be useful to also include out-of-transit spectra in the SP.
As a note of caution, we emphasize that neither the target signal nor any other nuisance
signal such as telluric lines or variations caused by stellar activity are included in the
model in spectral sections covered by SP regions.
If the SP regions are arranged such that specific wavelength intervals are
always excluded from the modeling, the model remains unconstrained for these
wavelength regions. In our mock setting (Fig.~\ref{fig:mock_emap}), the same spectral range is covered by
at most three SP regions in consecutive observations. This form of SP is, therefore, suitable
for fast moving, narrow signals, but not for the analysis of chemical species, such as many molecules, exhibiting
a dense population of features.   

\begin{table}[] 
    \centering
		\caption{Parameters of the planetary signal$^a$. 
	\label{tab:sysrem_mock}}
	\begin{tabular}{l c c c}
	\hline
	\hline
    \noalign{\smallskip}
	$k$ &  EW [m\AA]  & $\mu$ [\AA]   & $\sigma$ [$10^{-2}$ \AA] \\
    \noalign{\smallskip}
	\hline
    \noalign{\smallskip}
	\multicolumn{4}{c}{Without SP} \\
	1  &  $19.9\pm 1.3$ & $6565.058\pm 0.004$ & $6.7\pm 0.5$ \\
	2  &  $18.0\pm 1.0$ & $6565.060\pm 0.006$ & $6.9\pm 0.5$ \\
	3  &  $15.8\pm 1.3$ & $6565.059\pm 0.006$ & $6.8\pm 0.6$ \\
	5  &  $12.8\pm 1.2$ & $6565.059\pm 0.007$ & $6.8\pm 0.8$ \\
	10  &  $7.4\pm 1.2$ & $6565.057\pm 0.010$ & $6.0\pm 4.0$ \\
	    \noalign{\smallskip}
	\multicolumn{4}{c}{With SP} \\
	1  &  $20.2\pm 1.4$ & $6565.059\pm 0.004$ & $6.7\pm 0.4$ \\
	2  &  $20.1\pm 1.1$ & $6565.059\pm 0.004$ & $6.6\pm 0.4$ \\
	3  &  $20.0\pm 1.2$ & $6565.059\pm 0.004$ & $6.3\pm 1.8$ \\
	5  &  $19.8\pm 1.2$ & $6565.060\pm 0.006$ & $6.4\pm 0.5$ \\
	10  &  $20.4\pm 1.6$ & $6565.058\pm 0.005$ & $6.5\pm 1.8$ \\
    \noalign{\smallskip}
	\hline
	\end{tabular}
\tablefoot{$^a$ EW, central wavelength, and standard deviation of the recovered Gaussian mock planetary absorption signal as a function of the number $k$ of \sys\ iterations with and without SP.}
\end{table}

\subsubsection{Overfitting}
\label{sec:overfitting}
The \sys\ model has $M+N$ free parameters per iteration (Sect.~\ref{sec:sysrem}).
After $k$ \sys\ iterations, a total of $k\times(M+N)$ best-fit parameter values have been estimated.
Fitting a model with a large number of parameters can lead to overfitting, which means that the model
starts to reproduce not only the signal but also the random noise structure of the data.  
As a demonstration of the effect, we generated
30 mock observations \`a 200 data points, all independent realizations of a standard
normal distribution with zero mean, and applied \sys\ to these data, which of course contain no signal.

\begin{figure}[]
		\includegraphics[width=0.49\textwidth]{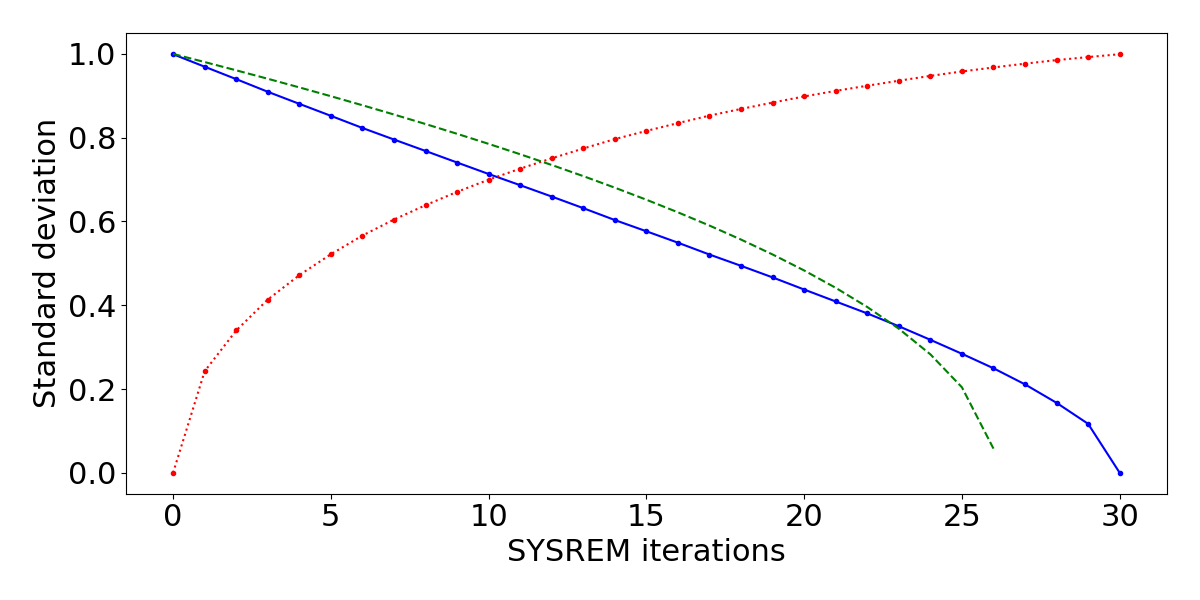}
	\caption{Standard deviation of residuals as a function of the number of \sys\ iteration in the application to white noise (solid blue),
	the approximation based on the degrees of freedom (green dotted), and the standard deviation of the cumulative model (dotted red).
	\label{fig:overfitting}}
\end{figure}

In Fig.~\ref{fig:overfitting}, we show the standard deviation of the residuals as a function
of the number, $k$, of \sys\ iterations
\begin{equation}
	\sigma_{\rm Res}(k) = \sqrt{\frac{1}{N \times M} \sum_{i=1}^N \sum_{j=1}^{M} \left(r^{(k)}_{i,j}\right)^2 } ,
\end{equation}
where $r^{(k)}_{i,j}$ denotes the residual after $k$ iterations.
The initial model, prior to the first iteration, is taken to be zero in all points, which is correct
by construction and any subsequent iteration overfits the data more severely. 
As the number of iterations increases, the standard deviation of the residuals decreases. Simultaneously, the standard deviation of the
model rises, which incorporates the noise (Fig.~\ref{fig:overfitting}).
Assuming that any of $k$ iterations removes $M+N$ degrees of freedom from
the sum of squares of the residuals, as is usual in linear regression problems \citep[e.g.,][]{Darlington2016},
a model for the evolution of the standard deviation may be written as 
\begin{equation}
	\sigma_{\rm Res}(k) \approx \sqrt{1 - k \, \frac{M+N}{M\times N}} ,
\end{equation}
which is a useful expression only as long as the argument of the square root remains positive.
As shown in Fig.~\ref{fig:overfitting}, this model approximately describes the empirical result
obtained for the standard deviation though the latter drop faster, which would yield a reduced
$\chi^2$ value smaller than one, again consistent with overfitting.  
 
Over a large range
of \sys\ iterations, the evolution in the standard deviation of residuals is approximately linear.
By carrying out sufficiently many of them, the residuals can be
diminished to essentially any level.
While overfitting may not be an actual problem in many applications, it is a problem if the
magnitude of the residuals is critical for the result as, for instance, when the residuals are used
to estimate upper limits.

\subsection{Transmission spectroscopy of \wa}
\label{sec:ts}
In the following, we apply the technique described in Sect.~\ref{sec:sysremTS} to
our \carm\ data of \wa. In particular, we study the \nad$_{1,2}$, H$\alpha$, Ca~IRT,
and \hei\ lines. 

\begin{figure}
		\includegraphics[width=0.49\textwidth]{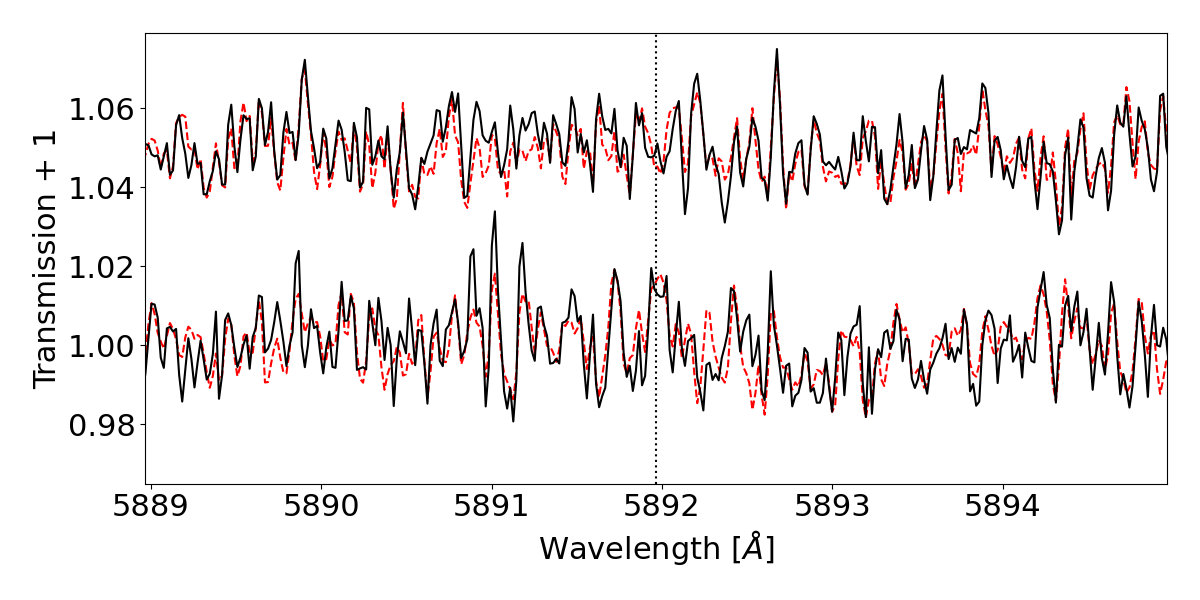}\\
	\includegraphics[width=0.49\textwidth]{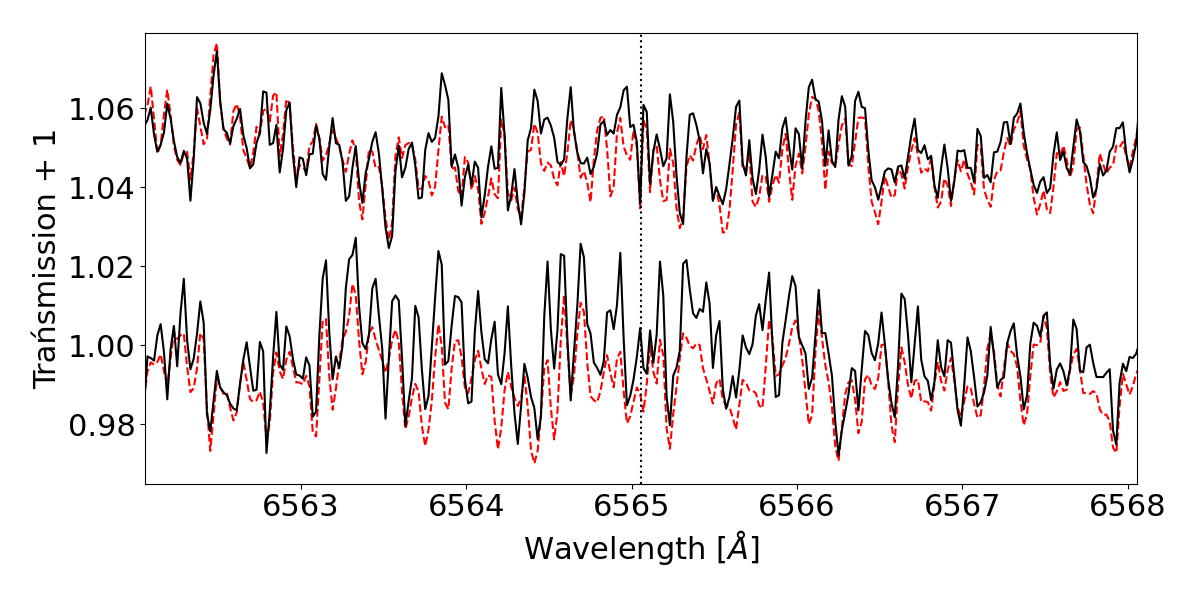}\\
	\includegraphics[width=0.49\textwidth]{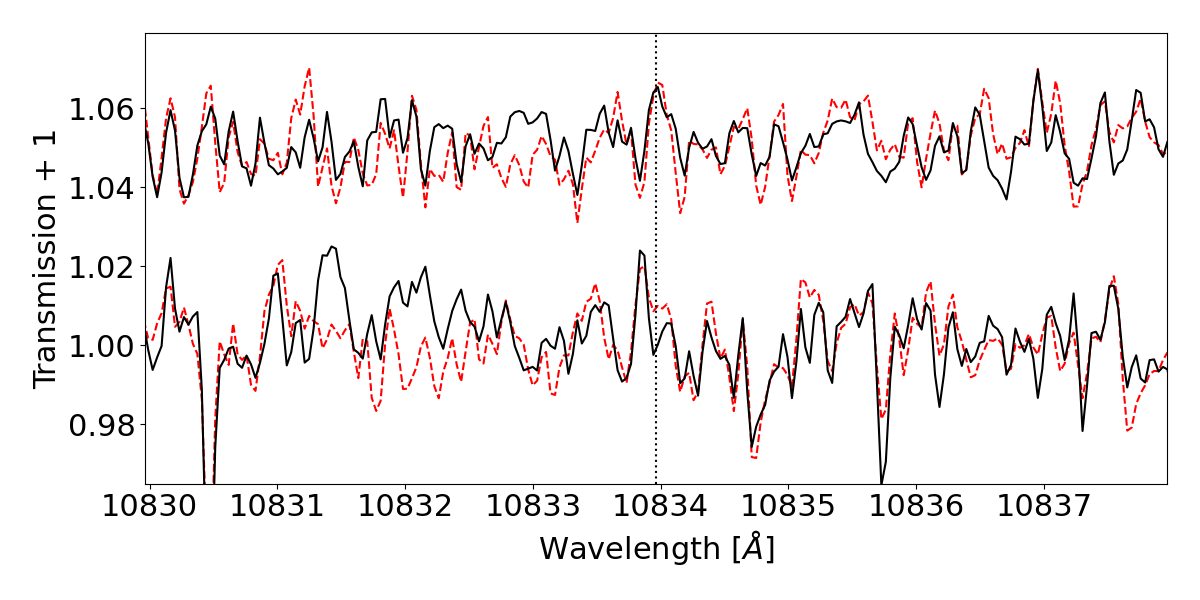}
	\caption{Transmission spectra of the \nad$_2$ (top), H$\alpha$ (middle), and \hei\ triplet lines
	(bottom) obtained using \sys\ (solid black) and a conventional approach (dashed red, Sect.~\ref{sec:comp_trad}).
	The night~2
	spectra are shifted by 0.05. The dotted vertical line indicates the nominal wavelength of the line
	(the center of the stronger doublet in the case of the \hei\ triplet).
	\label{fig:tss}}
\end{figure}

To obtain the transmission spectra in the PF, we evaluate the estimator $\mathcal{T}$ (Eq.~\ref{eq:estimator_T}),
after 1,2,3,5 and 10 \sys\ iterations, using the OF for the \sys\ input matrices. 
As in-transit spectra, we adopt the observations with the
running numbers $6-14$ for night~1 and $5-13$ for night~2 (see App.~\ref{sec:obslog}).
As examples, we show the thus obtained estimates of the transmission spectra for the \nad$_2$, \ha, and the \hei\ triplet lines
in Fig.~\ref{fig:tss} along with transmission spectra obtained using a more conventional methodology (Sect.~\ref{sec:comp_trad}).
Based on the
transmission spectra, we calculate the EW of the signal in regions with a width of one, two, and eight
times the expected FWHM of a narrow absorption signal (see Sect.~\ref{sec:tt}) both with and without applying SP.

To obtain the standard deviation of the EW estimate under the hypothesis that no planetary absorption signal
exists, we carry out a bootstrap analysis. To that end, we shuffle the residual spectra obtained after the application
of \sys\ in the OF by randomly (and with replacement) drawing rows from $\vec{R}^{(k)}_{\rm OF}$ to destroy the
temporal order of the series. We then proceed to transform into the PF, obtain the averaged in-transit spectrum, and
calculate the EW. The distribution of the EW is found by repetition of these steps and can be used to estimate the
significance of the actually found value or upper limits in the case of a non-detection.

\subsubsection{\ha\ line}

We first turn to the \ha\ line.
In Table~\ref{tab:hauls}, we report the derived $95$\,\% confidence level upper limits for the line
as a function of the number of \sys\ iterations and the width of the respective interval, centered on
the nominal wavelength of the \ha\ line. Signal protection was applied. The SP regions were centered on
the changing nominal orbital velocity of the \ha\ line, covering a range of
$\pm 15$~\kms, which roughly corresponds to the orbital RV shift per observation. This range covers three times
the FWHM of a phase-smeared, narrow planetary signal.   
Switching off SP only marginally affects the numbers.
The upper limit and standard deviation of the transmission spectrum decreases as the number of \sys\
iterations increases, which we attribute to overfitting (Sect.~\ref{sec:overfitting}).
To estimate the influence on the
upper limit, we take advantage of
the approximately linear evolution of the noise standard deviation observed in Sect.~\ref{sec:overfitting} by
fitting the evolution of upper limits and standard deviations as function of \sys\ iteration for all considered band widths 
with a linear model. We then evaluate this model at the hypothetical point of zero iterations.
The result is given in the bottom row of Table~\ref{tab:hauls}, where $0^*$ indicates the reconstruction, and we consider
this the best estimate of the upper limit.
The reconstructed values are similar to those obtained after the first iteration, indicating that
the outcome is not strongly affected by one iteration and that result may be used as an
appropriate estimate for the upper limit and standard deviation, if a single iteration is suitable to appropriately model
the observation in the first place (see Sect.~\ref{sec:ts_he} for a counterexample).   

\begin{table}[]
\centering
\caption{EW of \ha\ as a function of \sys\ iterations and absorption band widths$^a$.
\label{tab:hauls}}
\centering
\begin{tabular}{l l l l l | l l l l}
\hline\hline
\noalign{\smallskip}
   & \multicolumn{4}{c|}{Night 1} & \multicolumn{4}{c}{Night 2} \\
\# & \multicolumn{3}{c}{95\,\% UL [m\AA]} & $\sigma$ [\%] & \multicolumn{3}{c}{95\,\% UL [m\AA]} & $\sigma$ [\%] \\
   & 1 & 2 & 8 &  & 1 & 2 & 8 & \\  
\noalign{\smallskip}
\hline
\noalign{\smallskip}
 1 & 2.2 & 3.4 & 7.9 & 0.82 & 1.8 & 2.8 & 7.0 & 0.65 \\
 2 & 1.9 & 2.9 & 6.0 & 0.81 & 1.8 & 2.8 & 6.0 & 0.65 \\
 3 & 2.0 & 2.9 & 6.5 & 0.81 & 1.6 & 2.4 & 5.4 & 0.59 \\
 5 & 1.8 & 2.5 & 6.1 & 0.77 & 1.4 & 1.9 & 4.1 & 0.56 \\
10 & 1.5 & 2.2 & 4.8 & 0.65 & 1.2 & 1.6 & 3.3 & 0.47 \\
\noalign{\smallskip}
\hline
\noalign{\smallskip}
0$^*$ & 2.2 & 3.2 & 7.3 & 0.85 & 1.9 & 2.9 & 6.8 & 0.67 \\
\noalign{\smallskip}
\hline 
\end{tabular}
\tablefoot{$^a$ The $95$\,\% upper limits on the EW of an in-transit absorption signal as a function of \sys\ iterations (\#) for the \ha\ line in bands covering 1, 2, and 8 times the FWHM of a narrow line centered on the nominal wavelength in the planetary frame along with the standard deviation ($\sigma$) of the transmission spectrum and the reconstructed zeroth iteration values ($0^*$).}
\end{table}

\begin{table}
\centering
\caption{Same as Table~\ref{tab:hauls} but for the \nad$_1$ and \nad$_2$ and the \cairt\ lines, but reduced to the results for the first \sys\ iteration and the reconstructed zeroth iteration.
\label{tab:alluls}}
\begin{tabular}{l l l l l | l l l l}
\hline\hline
\noalign{\smallskip}
   & \multicolumn{4}{c|}{Night 1} & \multicolumn{4}{c}{Night 2} \\
\# & \multicolumn{3}{c}{95\,\% UL [m\AA]} & $\sigma$ [\%] & \multicolumn{3}{c}{95\,\% UL [m\AA]} & $\sigma$ [\%] \\
   & 1 & 2 & 8 &  & 1 & 2 & 8 & \\  
\noalign{\smallskip}
\hline
& \multicolumn{8}{c}{\nad$_1$} \\
1 & 1.3 & 2.0 & 4.6 & 0.76 & 1.4 & 2.1 & 5.0 & 0.64 \\
0$^*$ & 1.4 & 1.9 & 4.4 & 0.78 & 1.4 & 2.1 & 4.6 & 0.63 \\
\noalign{\smallskip}
\hline
\noalign{\smallskip}
& \multicolumn{8}{c}{\nad$_2$} \\
1 & 1.3 & 2.3 & 5.0 & 0.71 & 1.3 & 1.9 & 3.7 & 0.52 \\
0$^*$ & 1.4 & 2.4 & 5.4 & 0.72 & 1.3 & 1.8 & 3.8 & 0.54 \\
\noalign{\smallskip}
\hline
\noalign{\smallskip}
& \multicolumn{8}{c}{\cairt$_1$} \\
1 & 1.7 & 2.7 & 5.7 & 0.68 & 1.7 & 2.3 & 4.9 & 0.59 \\
0$^*$ & 1.8 & 2.8 & 6.1 & 0.70 & 1.8 & 2.4 & 5.3 & 0.59 \\
\noalign{\smallskip}
\hline
\noalign{\smallskip}
& \multicolumn{8}{c}{\cairt$_2$} \\
1 & 2.6 & 3.6 & 6.6 & 0.94 & 2.4 & 3.8 & 7.2 & 0.78 \\
0$^*$ & 2.8 & 3.9 & 7.3 & 0.95 & 2.6 & 4.0 & 7.7 & 0.81 \\
\noalign{\smallskip}
\hline
\noalign{\smallskip}
& \multicolumn{8}{c}{\cairt$_3$} \\
1 & 2.3 & 3.3 & 6.5 & 0.76 & 2.3 & 3.7 & 8.1 & 0.75 \\
0$^*$ & 2.4 & 3.4 & 6.7 & 0.76 & 2.2 & 3.6 & 8.3 & 0.77 \\
\noalign{\smallskip}
\hline
\end{tabular}
\end{table}

\begin{table}[]
\centering
\caption{Same as Table~\ref{tab:hauls} but without SP for the \hei\ lines.
\label{tab:heuls}}
\centering
\begin{tabular}{l l l l l | l l l l}
\hline\hline
\noalign{\smallskip}
   & \multicolumn{4}{c|}{Night 1} & \multicolumn{4}{c}{Night 2} \\
\# & \multicolumn{3}{c}{95\,\% UL [m\AA]} & $\sigma$ [\%] & \multicolumn{3}{c}{95\,\% UL [m\AA]} & $\sigma$ [\%] \\
   & 1 & 2 & 8 &  & 1 & 2 & 8 & \\  
\noalign{\smallskip}
\hline
\noalign{\smallskip}
 1 & 15.5 & 20.6 & 67.0 & 2.54 & 8.9 & 21.9 & 54.7 & 1.54 \\
 2 & 6.0 & 7.2 & 17.4 & 0.91 & 2.4 & 4.5 & 10.4 & 0.48 \\
 3 & 3.7 & 4.6 & 9.8 & 1.01 & 2.1 & 4.8 & 9.4 & 0.47 \\
 5 & 2.7 & 3.7 & 7.7 & 0.75 & 1.9 & 2.9 & 6.2 & 0.44 \\
10 & 2.2 & 3.3 & 7.4 & 0.54 & 1.4 & 2.4 & 4.6 & 0.43 \\
\noalign{\smallskip}
\hline
\noalign{\smallskip}
0$^*$ & 3.3 & 4.1 & 8.1 & 0.97 & 2.5 & 3.4 & 7.8 & 0.44 \\
\noalign{\smallskip}
\hline 
\end{tabular}
\end{table}

\subsubsection{\hei\ line}
\label{sec:ts_he}

\begin{figure}
		\includegraphics[width=0.49\textwidth]{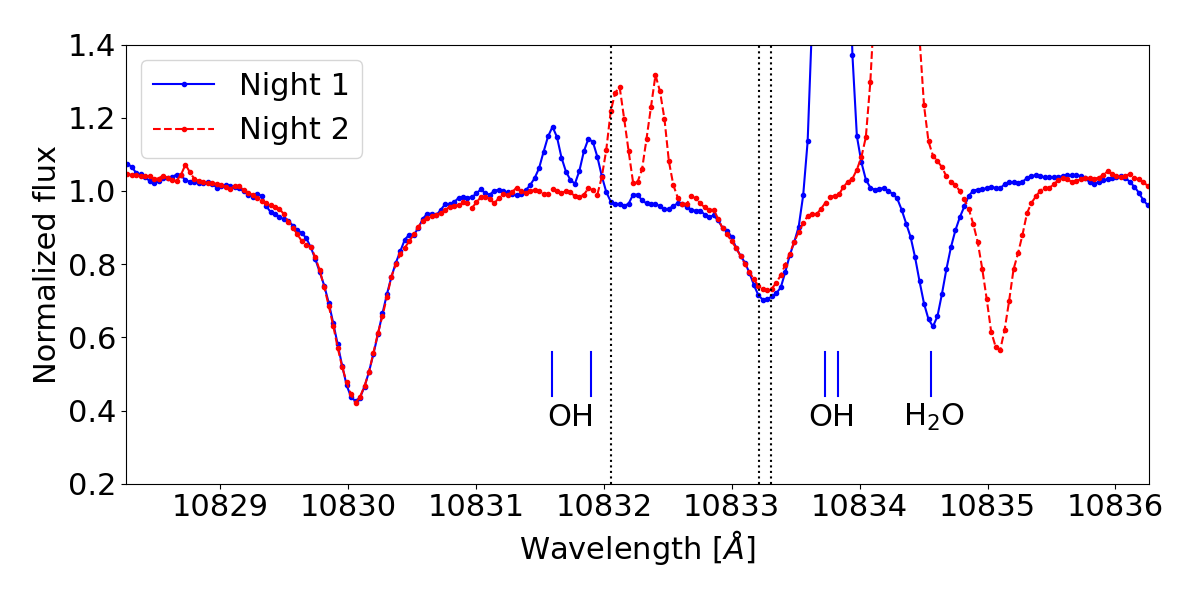}
	\caption{Average spectrum of \wa\ during night~1 and 2 in the SF before correction for telluric
	absorption lines. Dotted vertical lines indicate the nominal wavelengths of the lines of the
	\hei\ triplet. Marks at the bottom indicate the nominal wavelengths of the four OH emission lines
	and the strongest water line during night~1.   
	\label{fig:hemaster}}
\end{figure}

The \hei\ lines are strongly affected by variable telluric emission lines
\citep[e.g.,][]{Nortmann2018, Czesla2022}. This is also clearly visible in Fig.~\ref{fig:hemaster},
where we show the night-averaged spectra for nights~1 and 2 around the \hei\ lines in the SF. The spectra
show strong OH doublets close to the \hei\ triplet \citep{Czesla2022}, which would reach peak heights of about
three and five in units of normalized flux for nights~1 and 2.
Additionally, a prominent water absorption line
is present. All telluric lines shift in the figure owing to the choice of reference system. 
Curiously, the stronger blended doublet of the \hei\ triplet appears to be deeper in night~1, which is discussed further
in Sect.~\ref{sec:internight_he}. 

\begin{figure}[]
		\includegraphics[width=0.49\textwidth]{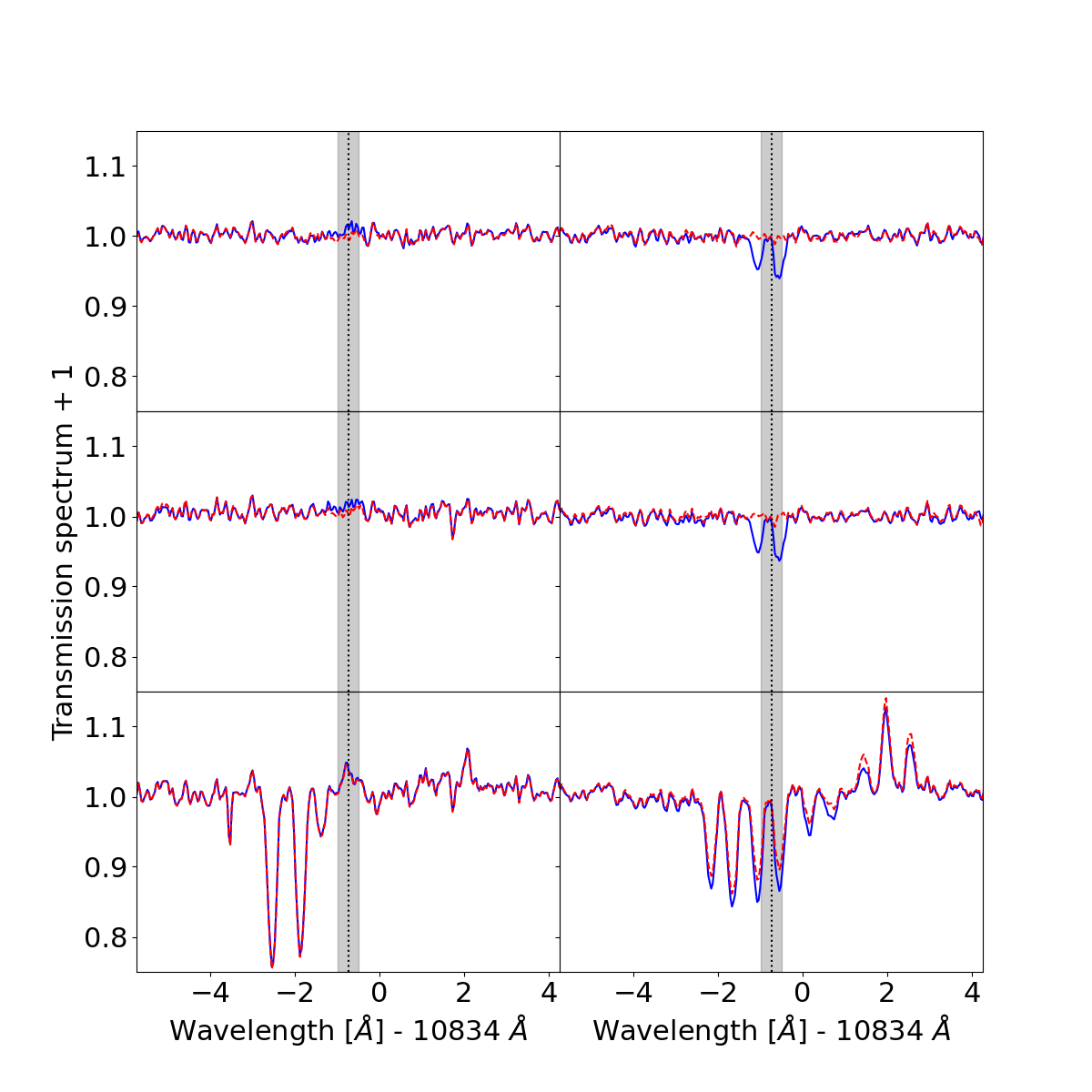}
	\caption{Transmission spectra after one (bottom row), three (middle row),
	and five (top row) \sys\ iterations for night~1 (left column) and night~2
	(right column) with SP (solid blue) and without (dashed red). The gray shade
	indicates the width of the SP region.
	\label{fig:hensys}}
\end{figure}

In Fig.~\ref{fig:hensys}, we show the transmission spectrum in the PF as obtained
after one, three, and five \sys\ iterations both with and without applying SP.
After one iteration, the impact of the OH emission lines is clearly seen, as they shift through the
spectrum in the PF. After three and five iterations, the relics
are removed in the night~1 transmission spectra. For the night~2 data, we find a
marked difference between the transmission spectra obtained with and without SP.
In particular, the former show
strong relics (in apparent absorption) of the OH lines within a range compatible with the width of the adopted SP region,
while the latter do not. This is clearly a consequence of the SP not selectively shielding
the planetary signal but also the underlying OH lines, which are not correctly interpolated
by the \sys\ model.

In Table~\ref{tab:heuls}, we show the bootstrap-derived upper limits for possible
absorption in the \hei\ lines, obtained without applying SP. After one and possibly also
after two iterations, the values strongly decrease indicating that the overfitting regime
had not yet been reached and real spectral structure such as the OH lines
were still incorporated into the model.
Therefore, we adopted only the values obtained after three or more iterations in our
linear reconstruction (Sect.~\ref{sec:ts}).

\subsubsection{Comparison with conventional approach}
\label{sec:comp_trad}

The conventional approach to obtain planetary transmission spectra of atomic lines does not involve \sys,
but relies on a comparison of spectra taken during the event with those before and after \citep[e.g.,][]{Wyttenbach2015}.
Typically, a reference out-of-transit spectrum, $f_{\rm OOT}$, is obtained by averaging
spectra taken before and after the transit, which do supposedly not contain the planetary signal.
Residual spectra are then obtained by dividing the observed
spectra by $f_{\rm OOT}$ in the SF or BF. These residual spectra are subsequently shifted into the PF. Averaging
them over some time interval of interest such as the transit yields the associated transmission spectrum.
Regions strongly contaminated by stellar, telluric, or instrumental signals can be masked in the process,
which of course leads to a loss of the masked signal. 

To compare our results, we also
show transmission spectra obtained by this conventional approach in Fig.~\ref{fig:tss}. In the case of the
\hei\ lines, masking regions with a total width of four times the instrumental FWHM were placed on the four offending OH
lines at vacuum rest wavelengths of 10\,832.412~\AA, 10\,832.103~\AA, 10\,834.338~\AA, and 10\,834.241~\AA\ \citep{Oliva2015, Czesla2022},
with the final mask being the union of the individual ones.

The resulting transmission spectra are very similar. In most instances, structure on the scale of the instrumental
resolution is reproduced. This indicates that the conventional reference spectrum and the model constructed by
\sys\ bear strong similarity. For the \nad$_2$ line (top panel in Fig.~\ref{fig:tss}), we did not mask the telluric emission. 
The fast RV change of the planetary signal obviously helps to smear this out in the conventional analysis. For the \ha\ line
transmission spectrum, some difference on a larger scale is visible particularly for the night~1 transmission spectra. We speculate
that this is caused by variation in the broad core region of the \ha\ line during the transit, which is not accounted for in
the conventional approach, but incorporated into the \sys\ model. Also the transmission spectra of the \hei\ lines show
rather insignificant differences given the noise. This clearly shows that the modeling results obtained by \sys, here after
three iterations, are at least on a par with that produced by complete masking of the region.

\subsection{\ha, \cairt, and \hei\ line variation in the stellar frame}
\label{sec:internight}

Our night~1 and 2 observing runs both last for approximately six hours, which
are separated by 335~days. Therefore, stellar spectral changes
on the hour and year timescale can be examined. Variability in the stellar spectrum
of \wa\
may potentially be caused by magnetic activity or changes in absorption by
circumstellar material such as a giant atmosphere or
torus-like structure  \citep[][]{Fossati2013, Haswell2012, Debrecht2018}.The \ha\ and \hei\ triplet lines are potentially
sensitive to both of these effects.

\subsubsection{The \ha\ and \cairt\ lines}

\begin{figure}
			\includegraphics[width=0.49\textwidth]{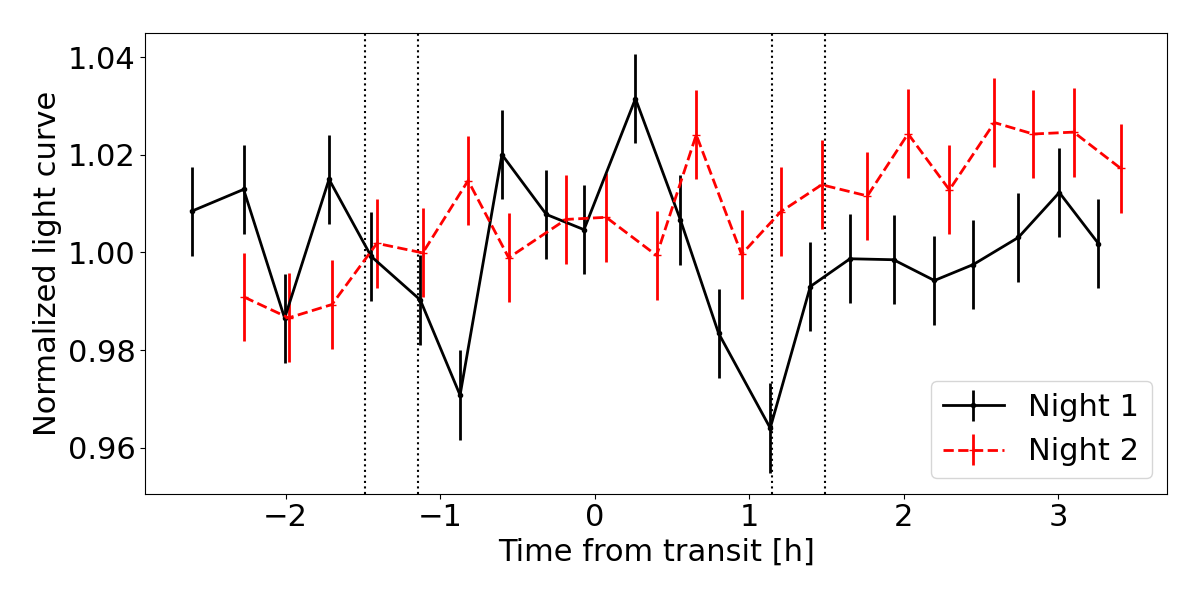}\\
	\includegraphics[width=0.49\textwidth]{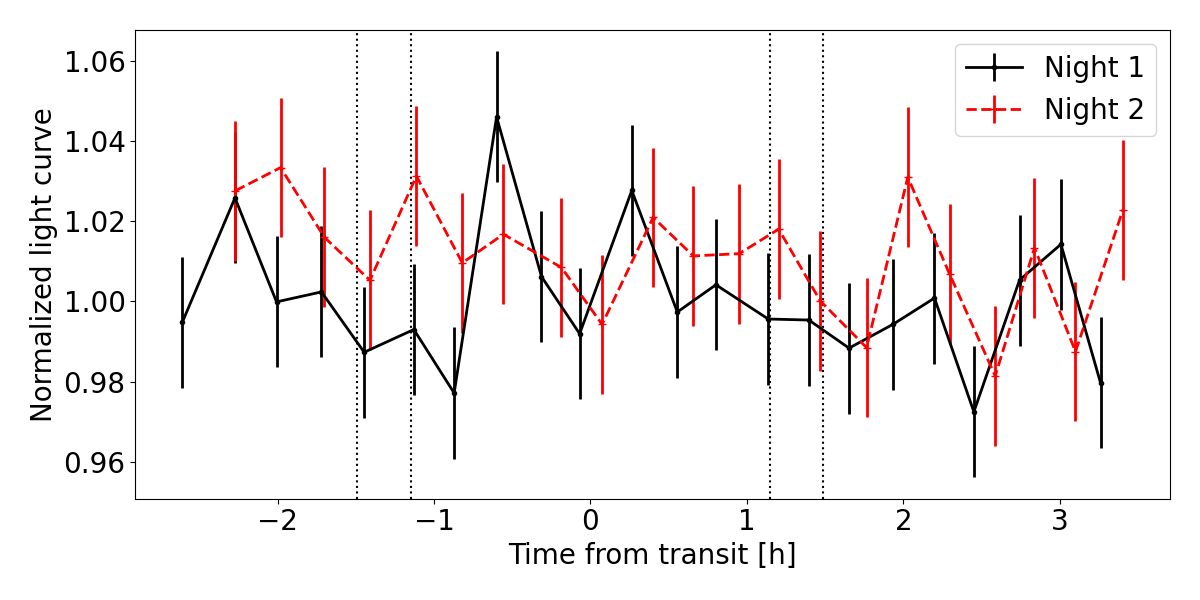}
	\caption{Light curves of the \ha\ and \cairt\ lines in the stellar frame. 
	Top: \ha\ light curves (1.5~\AA\ passband) normalized by the mean of
	the night~1 light curve. Bottom: As top but for $0.5$~\AA\ passband centered on bluest component of
	the \cairt.
	\label{fig:halcorig}}
\end{figure}

In Fig.~\ref{fig:halcorig}, we show the \ha\ and \cairt\ line light curves during the two observing nights.
For the \ha\ line, the light curves were obtained by integrating
the flux density in a 1.5~\AA\ wide band centered on the nominal wavelength of the \ha\ line in the SF. The light curve
was normalized by the mean of integrals of the flux density in two reference bands ($6550-6555$~\AA\ and $6578-6583$~\AA), and,
finally, the night~1 and 2 light curves were divided by the mean of the night~1 light curve. For the \cairt\ line, a $0.5$~\AA\ wide
band was adopted. 
Uncertainties were estimated from the data using the $\beta\sigma$ estimator \citep[][]{Czesla2018}.
No clear flaring can be discerned in the light curves.
During night~1, the \ha\ light curve shows more short-term variation primarily before the egress and conspicuous
drops close to the in- and egress phases, which are,
however, not reproduced in night~2. This shape is independent of whether the telluric correction is used.
As also the night~1 S/N ratio is lowest before egress (Fig.~\ref{fig:airmass}), we consider
it unlikely that the variation is astrophysical in origin. 
The night~2 light curve shows
a systematic rise, indicative of
a continuous fill-in of the \ha\ line center, which is not observed during night~1.
We approximated the \ha\ light curves by a linear model and used
a bootstrap analysis to show that the null hypothesis of a constant light curve
can confidently be rejected (p-value $5\times 10^{-5}$) for the night~2 light curve, while
the night~1 light curve is consistent with being constant. If anything, the corresponding \cairt\
light curve shows a decline, which is, however, not highly significant (p-value $4.6$\,\%).

\begin{figure}[]
		\includegraphics[width=0.49\textwidth]{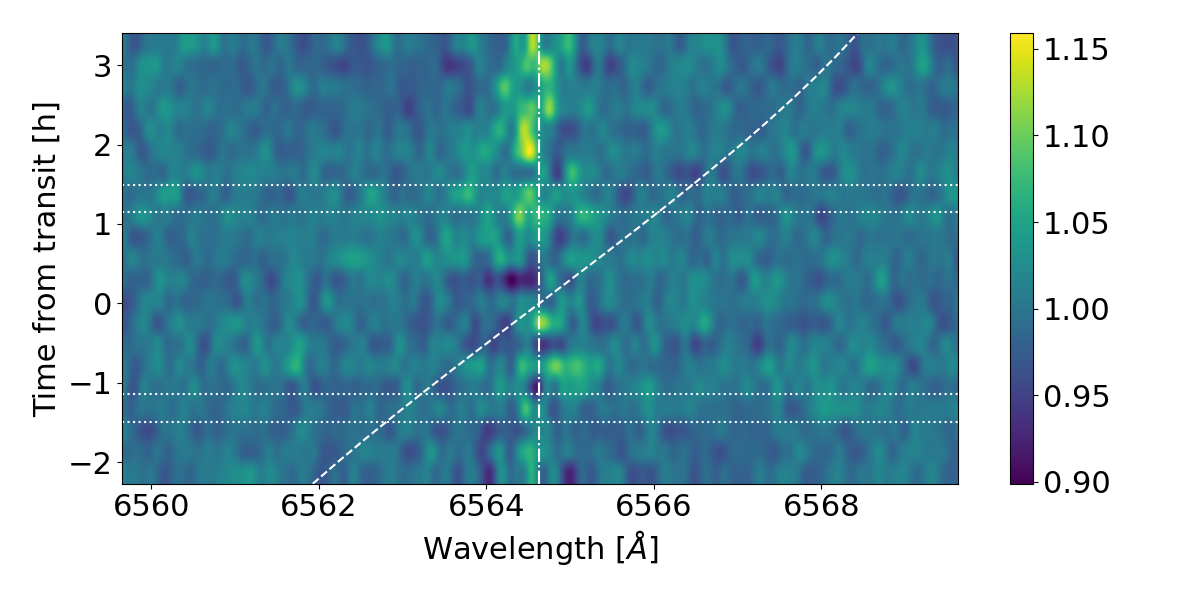}
	\caption{Ratio of night~2 and 1 spectra in the SF as a function of time from transit center. Horizontal
	dotted lines indicate the first to fourth contact, the dash-dotted vertical line denotes the
	nominal position of the \ha\ line, and the dashed diagonal line shows the orbital velocity track
	of the planet.
	\label{fig:ha_ratmap}}
\end{figure}

\begin{figure}[]
		\includegraphics[width=0.49\textwidth]{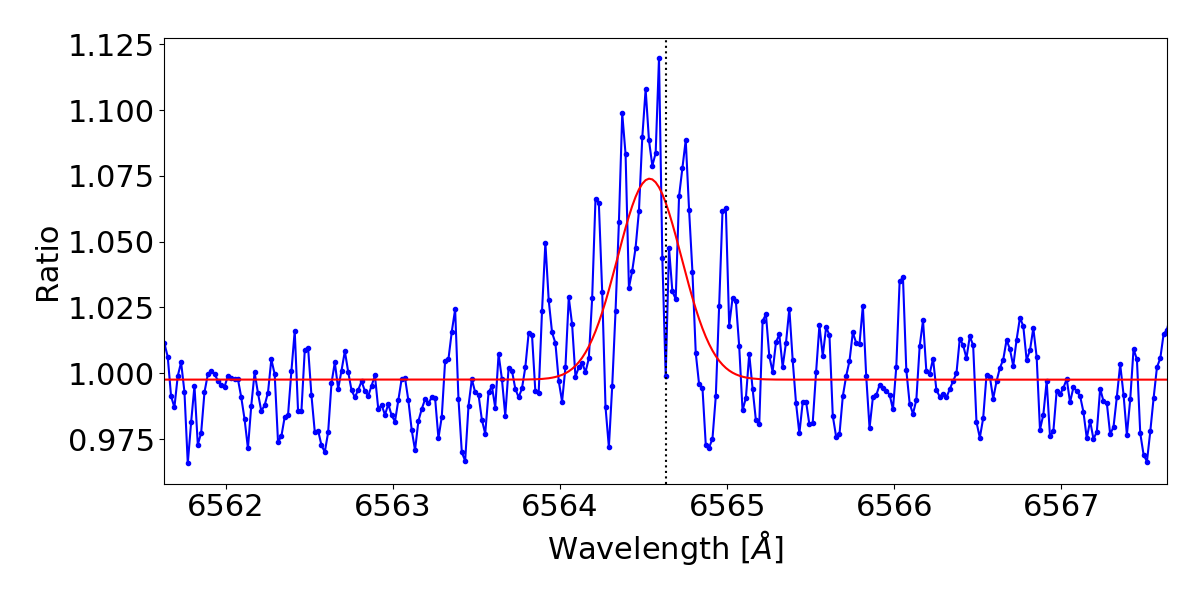}
	\caption{Average post-transit ratio of night~2 and night~1 spectra around the \ha\ line (blue)
	along with best-fit Gaussian model (red solid). 
	\label{fig:ha_postrat}}
\end{figure}

To get a better view of the actual spectral changes, we now compare the spectra of night~1 and 2 directly.
In night~1, we obtained one more spectrum and the observing run commenced at slightly earlier orbital
phase (e.g., Fig.~\ref{fig:airmass}). To compare the spectral evolution between the two nights at similar
orbital phases, we, therefore, discard the first spectrum of night~1 and compare the remaining spectra.
The maximum thus obtained orbital phase offset between the night~1 and 2 observations is $5.7\times 10^{-3}$
or $9$~min. Dividing the night~2 by the night~1 spectra, we obtain the ratios shown in Fig.~\ref{fig:ha_ratmap}.
The most pronounced differences can be distinguished in the vicinity of the core of the \ha\ line in the SF
after the nominal fourth contact, which provides a useful but not necessarily physically meaningful temporal
subdivision of the time series in this context. The ratios indicate a fill-in of the \ha\ line core in night~2
compared to night~1, consistent with the light curve (Fig.~\ref{fig:halcorig}). In Fig.~\ref{fig:ha_postrat},
we show the mean post-transit spectral ratio, which clearly displays the night~2 fill-in. A Gaussian fit
of the feature yields an EW of $37 \pm 7$~m\AA, a FWHM of $0.45\pm 0.13$~\AA\ or $21\pm 6$~\kms, and a RV shift
of $-4.4\pm 1.5$~\kms.

A possible explanation for the differences in the core of the \ha\ line may arise from a change in the level
of stellar magnetic activity between night~1 and night~2 as well as during the night~2 observing run itself.
Examples of \ha\ line variability tantamount to our observation have been reported for 
M~dwarfs, which show comparable and larger long-term variations in the \ha\ EW \citep{Fuhrmeister2023}.
Rotational modulation in the \ha\ line with an EW amplitude of about $\pm 20$~m\AA\ was also observed in
the active K-dwarf HD~189733 \citep[][]{Kohl2018} and the fast-rotating, active G2 dwarf V~889~Her shows an amplitude
of about $\pm 50$~m\AA\ \citep{Frasca2010}. Yet, rotational modulation appears to be an unlikely cause for the
observed \ha\ line variations as \wa\ shows almost no spectroscopic rotational broadening, which is required to produce
rotational modulation irrespective of projection effects. Also the \cairt\ line provides no evidence for
a change in activity.

 A second conceivable explanation may be \ha\ line contamination from
the M-dwarf binary companion, however, we argue that this is unlikely.
Firstly, the M~dwarf binary is significantly fainter than the primary. 
\citet{Bechter2014} derived a spectral type of M3V for them, but the Gaia \citep{Gaia2023} upward revision of the
distance estimate (Table~\ref{tab:props}) suggests a slightly earlier spectral type
of M1V and an effective temperature of about $3600$~K \citep{Kraus2007}. We
consulted synthetic PHOENIX spectra calculated by \citet{Husser2013} to compare the flux densities of the primary and the secondary.
In the core of the \ha\ line, we find a ratio of $33$, and a ratio of $90$ in the close-by pseudo-continuum.
The observed contrast is further enhanced by
their separation of $1''$, which puts the binary beyond the acceptance angle of the $1.5''$ diameter fiber feeding
CARMENES \citep{Quirrenbach2018}. Although the seeing disks may show some overlap with the fiber pointed
at the primary, M1V type dwarfs show essentially no photospheric \ha\ line.
As activity and contamination appear unlikely, the variation probably originates in the \wa\ system itself.

A third explanation for the observed differences may be artifacts
from the data taking or reduction. As \wa\ is a relatively
faint target, the S/N of individual spectra remains moderate (Fig.~\ref{fig:airmass}). During night~1, the S/N is systemically
lower in the first half of the night. In such cases, and in particular in the deep trough of the \ha\ line, the relative
contribution of the background increases, so that small errors in its estimation may produce spurious changes in the line depth.
However, neither the light curve of the \ha\ nor the \cairt\ lines (Fig.~\ref{fig:halcorig})
show structure reminiscent of the night~1 step in S/N. Moreover, the observed systematic evolution in the line profile presumably took place
during night~2 when the S/N and, thus, signal level were more stable. While this does not rule out such a scenario, it does
not appear a likely explanation. Finally, the variation may be caused by variable atmospheric absorption further discussed in Sect.~\ref{sec:abshettorus}. 

\subsubsection{The \hei\ lines}
\label{sec:internight_he}

For the \hei\ triplet lines, changes in the SF are more challenging to quantify because of the strong telluric
contamination as demonstrated by the night-averaged \hei\ spectra of \wa\ shown in Fig.~\ref{fig:hemaster}.
Nonetheless, the same figures readily demonstrates that also the \hei\ line core in the night~2 spectrum appears to show a fill-in with respect to
the night~1 spectrum. We argue that this is likely not spurious, because in night~1, the strong OH line doublet located redward
of the \hei\ lines is much closer to the actual \hei\ line. If anything, it tends to add contaminating flux in the \hei\ core and
not subtract any. Likewise, any potential water-line contamination would tend to be deeper in night~2 (Fig.~\ref{fig:hemaster}) besides
from taken care of by the telluric correction.
The difference amounts to about $10$~m\AA\ at slightly above instrumental width.
Although we consider an actual change in the \hei\ line core likely, which shows the same trend as that observed in the \ha\ line core,
the numbers are unreliable owing to the contamination.

\subsection{Interstellar absorption in the \nad\ lines}
\label{sec:na_ism}
\begin{figure}
		\includegraphics[width=0.49\textwidth]{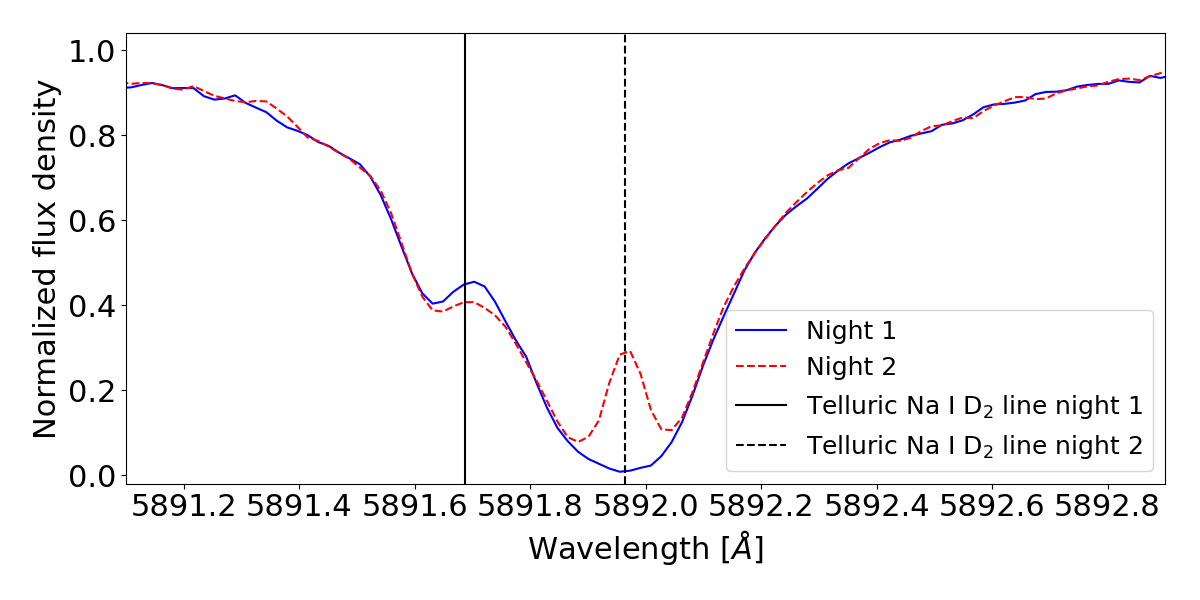}
	\caption{Average \nad$_2$ lines of \wa\ in nights 1 and 2. The average wavelengths of
	the telluric line are indicated by vertical lines. 
	\label{fig:natell}}
\end{figure}

The \nad\ lines of \wa\ are affected by interstellar and possibly circumstellar
absorption \citep{Fossati2013}. Additionally, our \carm\ observations show
telluric emission. In Fig.~\ref{fig:natell} the average spectra of the \nad$_2$
line of \wa\ during night~1 and 2 are shown. The impact of
telluric emission is particularly pronounced during night~2, where it occurs close to the
line core.
Clearly, the lower envelope of the two average spectra shown in Fig.~\ref{fig:natell}, which
we call the minimum master spectrum, $f_{\rm mima}(\lambda)$,
represents a better approximation of the true stellar line profile than any of the individual
spectra. 

\begin{figure}
		\includegraphics[width=0.49\textwidth]{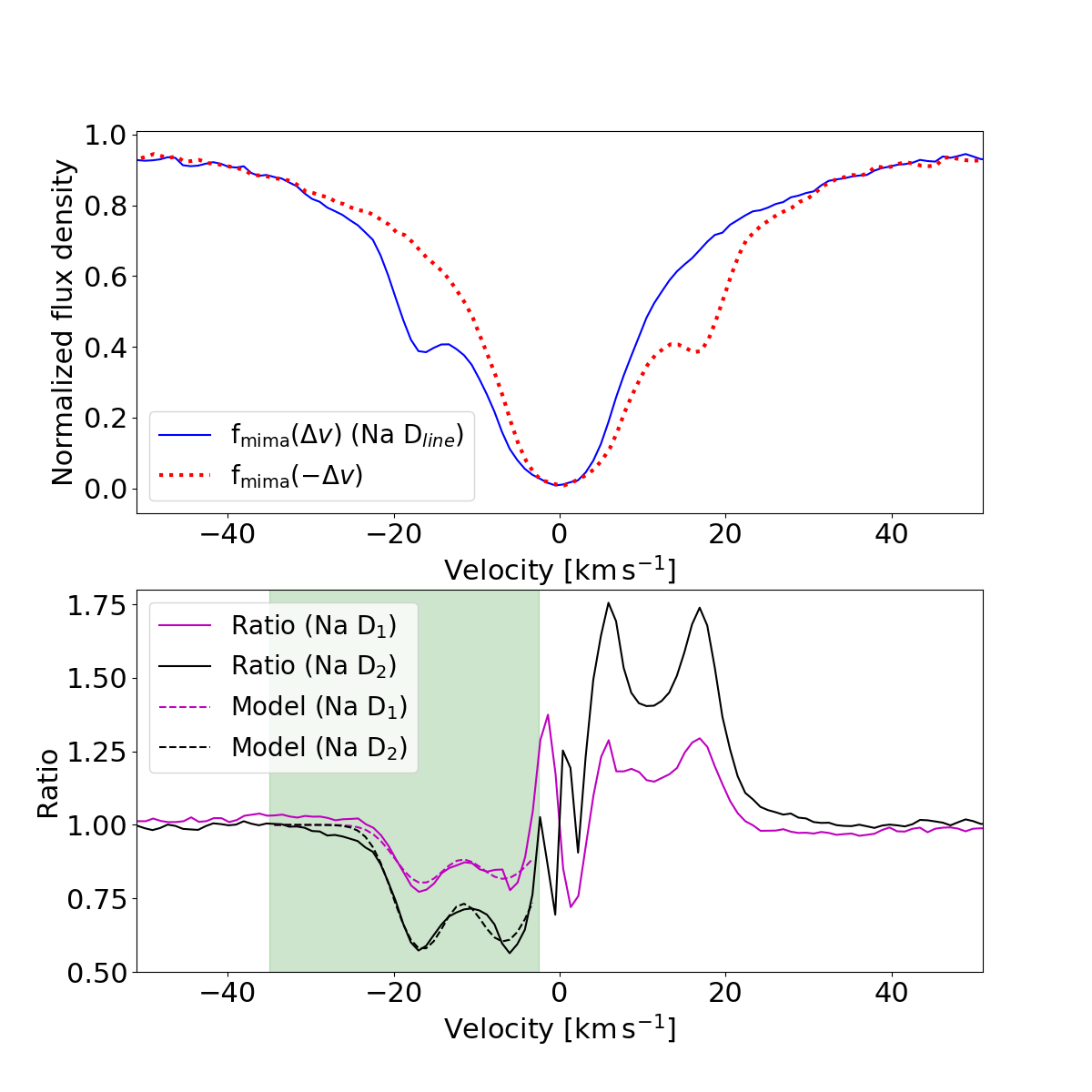}
	\caption{Properties of the \nad\ lines of \wa.
	Top panel: \nad$_2$ line of \wa\ (solid blue) along with spectrum reflected at
	nominal wavelength of the line (red dashed). Bottom panel: Ratio between original and
	reflected spectrum for \nad$_1$ and \nad$_2$ lines (solid magenta and black) along with
	best-fit model (dashed magenta and black). The green shade indicates the considered
	fit range.
	\label{fig:naism}}
\end{figure}

By investigating the \ion{Ca}{ii} H and K as well as the \nad\ lines of similarly distant early-type stars
at lines of sight close to \wa, \citet{Fossati2013}
identified at least two discernible absorption components attributable to interstellar material,
which are mainly blue shifted with regard to the spectrum of \wa.
To verify and examine the presence of interstellar absorption components in our spectra
of \wa, we used the minimum master spectrum of the \nad\ lines obtained above and assumed that the
stellar \nad\ lines are intrinsically symmetric. This premise is backed by synthetic PHOENIX spectra
\citep{Husser2013}, but factors such as stellar activity are not accounted for in these models.
We 
reflected the minimum master spectrum at the nominal wavelengths of the \nad\ lines (Fig.~\ref{fig:naism}, upper panel)
and used the reflection as an estimate of the unabsorbed stellar spectrum.
The lower panel of Fig.~\ref{fig:naism}
shows the ratio of the original and reflected minimum master spectrum, which clearly shows the effect of interstellar
absorption. This technique remains insensitive to absorption
in the line center and symmetric absorption in both wings of the \nad\ lines. 

We modeled the absorption in the blue flank between
velocity offsets of $-2.5$~\kms\ and $-35$~\kms\ (indicated
by shades in the figure) using a two-component model for the wavelength-dependent
optical depth, $\tau(\lambda)$, of the absorbers. The latter are parameterized as \citep[e.g.,][]{Axner2004} 
\begin{align}
	\tau(\lambda) = \sum_{l,j} N_l \alpha_j \lambda^2 \left(\frac{e^2}{4\epsilon_0 m_e c^2}\right)  \phi_l(\lambda-\lambda_{0,j}) \; .
	\label{eq:tau}
\end{align}
Here, the index $l \in \{{\rm r,b}\}$ denotes the two Gaussian components, which we dub blue (b) and red (r) to be consistent with
\citet{Fossati2013}, and the index $j \in {1,2}$ the individual sodium D$_1$ and D$_2$
lines with reference wavelengths $\lambda_{0,j}$,
$N_j$ is the column density of absorbing particles, $\alpha_j$ is the oscillator strength of the respective line, 
$e$ is the electric charge, $\epsilon_0$ is the vacuum permittivity, $m_e$ denotes the electron mass,
and $c$ is the speed of light. The profile functions $\phi_l$ are assumed to be Gaussian.
The free parameters of our model are the column densities, $N_l$, the widths of the Gaussian profile functions,
and their offset from the line centers. The fit is carried out for both lines simultaneously.

In Table~\ref{tab:naism}, we give our best-fit parameters with error derived using the jackknife method
\citep[e.g.,][]{Efron1981} along with the values reported by \citet{Fossati2013} for the line of sight
toward HD~257926, a star at a sky-projected separation angle of about $16'$ and, therefore, with a similar line of sight. 
The EW and column density derived for the blue absorption component are consistent
within the error margin. Our best-fit values for the FWHM and RV are somewhat smaller, which may be
attributed to a true difference in absorption between the lines of sights of \wa\ and HD~257926, or plausibly,
our continuum estimation method of reflecting the spectrum. The latter can also explain the differences
in the value determined for the red component, which we find to be considerably weaker than
reported by \citet{Fossati2013}. This component is located closer to the \nad\ line center and, thus, the
reflection axis of the spectrum, which explains our likely spurious finding of a weaker component.  

\begin{table}[]
    \centering
		\caption{Best-fit parameters of \nad\ ISM absorption components in \wa\ and values derived by
	\citet{Fossati2013}.
	\label{tab:naism}}
	\begin{tabular}{l c c l}
	\hline\hline
    \noalign{\smallskip}
	Parameter & Value & Literature$^a$ & Unit \\
    \noalign{\smallskip}
	\hline
    \noalign{\smallskip}
	RV$_{\rm b}$ $^b$ & $2.75\pm 0.16$ & $4.2 \pm 0.2$ & \kms \\
	RV$_{\rm r}$ $^b$ & $12.7\pm 0.7$ & $16.9\pm 0.1$ & \kms \\
	EW$_{\rm b}$ $^c$ & $72\pm 7$ & $79\pm 4$ & m\AA \\
	EW$_{\rm r}$ $^c$ & $90\pm 5$ & $212 \pm 3$ & m\AA \\
	$\log{\mbox{N}_{\rm b}}$ & $11.56\pm 0.04$ & $11.6\pm 0.1$ & cm$^{-2}$ \\
	$\log{\mbox{N}_{\rm r}}$ & $11.657\pm 0.023$ & $12.8\pm 0.1$ & cm$^{-2}$ \\
	FWHM$_{\rm b}$ $^c$ & $7.0 \pm 0.5$ & $10.5\pm 0.5$ & \kms \\
	FWHM$_{\rm r}$ $^c$ & $8.1 \pm 0.7$ & $10.1\pm 0.1$ & \kms \\
    \noalign{\smallskip}
	\hline
	\end{tabular}
\tablefoot{$^{(a)}$ From Table~2 of \citet{Fossati2013}; $^{(b)}$ In the barycentric system; $^{(c)}$ For the \nad$_2$ line.
}
\end{table}

\section{Discussion}

\subsection{\sys\ for individual line transmission spectroscopy}

In Sect.~\ref{sec:comp_trad}, we already showed that comparable results can be obtained
with \sys, used as described in Sect.~\ref{sec:sysremTS}, and a more conventional approach
to transmission spectroscopy at least in the case of our data. In the following, we compare the approaches
on a more conceptional level. 

In both approaches,
the handling of the residual spectra is identical once obtained, but
the construction of the reference differs.
In Sect.~\ref{sec:sysremTS}, the \sys\ model assumes a role comparable to the
out-of-transit reference spectrum by which the spectra are divided to
produce residuals in the conventional approach.
If used as described in Sect.~\ref{sec:sysremTS}, at least
two differences between these models stand out: (i) the \sys\ algorithm can incorporate the information of all spectra
into the model, as opposed to being limited to out-of-transit spectra and
(ii) for each spectrum an individual reference spectrum is constructed by \sys.

One advantage of using all spectra in the modeling is that the model has better leverage on approximately stationary though time-dependent
effects such as telluric absorption and emission lines as well as stellar activity. A case in point is our analysis of the transmission
signal of the \hei\ signal (Sect.~\ref{sec:ts_he}), which demonstrates that the strong OH lines are handled satisfactorily after
a few \sys\ iterations. Likewise, we found that the correction of telluric absorption lines was not critical for our results.
However, the model also gains leverage in the actual signal of interest,
which, at least for isolated spectral lines, may be countered by shielding the signal using
signal protection (Sect.~\ref{sec:signal_protection}).
The application of \sys\ is advantageous in situations where time-variable effects such as telluric emission
lines are strong, which complicates the construction of a suitable reference spectrum. Notably, useful results could
be obtained in our case without carrying out a telluric correction or the masking of other contaminants.   

\subsection{Comparison to previous tentative detections or non-detections}

\citet{Kreidberg2018} report a non-detection of \hei\ absorption in \wa. In particular, they
find the transit depth to be (insignificantly) elevated by $(49 \pm 143) \times 10^{-6}$,
in a 70~\AA\ wide band centered on the nominal wavelength of the \hei\ triplet. This
corresponds to an EW of $4\pm 10$~m\AA, which is consistent with our more stringent upper limits
(Table~\ref{tab:heuls}). 

\citet{Burton2015} find a tentative detection of absorption in the \nad\ lines at a level of $0.15\pm 0.03$~\%
across 2~\AA\ wide intervals, which corresponds to an EW of $3\pm 0.6$~m\AA\ per line, assuming that both
lines contribute equally. Given that $2$~\AA\ correspond to about 10 times the FWHM of a narrow signal in 
the \nad\ lines at our effective spectral resolution, the result is consistent with our upper limit (Table~\ref{tab:alluls})
insofar as we cannot rule out a broad signal with this EW. However, if the signal EW is more concentrated on the
line core region, we can give more stringent upper limits.  

\subsection{Absorption by a heterogeneous torus}
\label{sec:abshettorus}

In their analysis of the \nad\ and \ha\ lines of \wab,
\citet{Jensen2018} find excess absorption at levels of $52.6$~m\AA\ and $64.9$~m\AA,
respectively, measured across 2~\AA\ wide regions. The signals reach depths of about $6$\,\% in the \ha\ and $3$\,\% in the
\nad\ lines and both signals are blueshifted by about $-9$~\kms\ in the SF, though the authors caution that
their methodology may not be well suited to provide accurate atmospheric shifts. 
The width of the signal is consistent with their spectral resolution of $15\,000$, that is, around $20$~\kms.
The absorption signals studied by \citet{Jensen2018}
appear to be modulated on the timescale of the planetary orbital period. According to these authors,
this can be understood in the light of evidence for a possibly asymmetric circumstellar torus or disk-like structure
in the \wa\ system, ultimately formed by material originating from the planet \citep[e.g.,][]{Lai2010, Li2010, Haswell2012}.

\begin{figure}[]
		\includegraphics[width=0.49\textwidth]{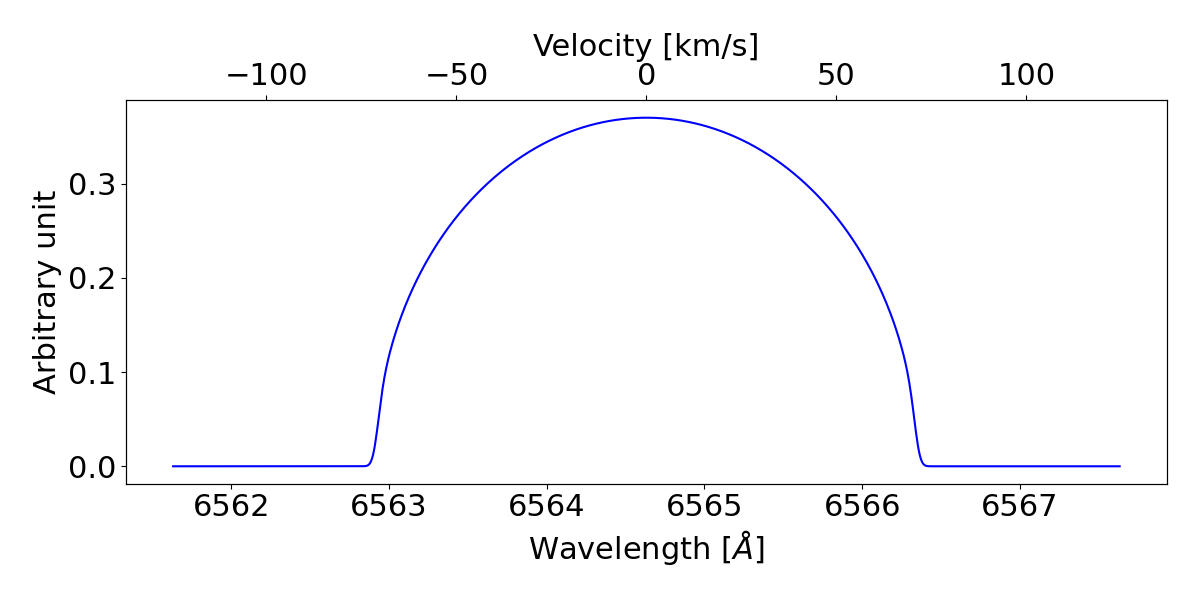}
	\caption{Hypothetical \ha\ absorption line profile caused by a homogeneous torus at the planetary orbital separation,
	covering the entire stellar disk.
	\label{fig:hatorus}}
\end{figure}

In our analysis,
we found no evidence for \ha\ absorbing material with a steady-state velocity structure,
moving along with the planetary body in the PF (Table~\ref{tab:hauls}).
If hypothetical excess absorption were caused by an extended structure producing a very prolonged transit,
as suggested by \citet{Jensen2018} for \wa\
and recently observed in the case of \hpb\ 
\citep[][]{Zhang2023}, our temporal baseline may not be long enough to cover the event in its entirety.
However, while this is also the case for the CARMENES data of \hpb\ analyzed by \citet{Czesla2022},
significant variation in the \ha\ and \hei\ lines is detected nonetheless. Also the UV transits studied by
\citet{Fossati2010b, Haswell2012, Nichols2015}, despite showing evidence for prolonged and potentially variable UV transits,
are consistent with the bulk of absorption occurring during the optical transit. A signal may of course still
be present in \wa, but too weak to be detected by us. 

The temporal evolution of the \ha\ line in the SF
is different in night~1 and night~2. In particular, a slightly blue-shifted fill-in with an EW of about $37$~m\AA\ develops
toward the end of our night~2 run. The signal width of about $21$~\kms\ is clearly above our instrumental resolution.
The width, shift, and strength of this signal are compatible with that observed by
\citet{Jensen2018}, accounting for the differences in spectral resolution. However, our signal was extracted by
comparing spectra obtained during the same planetary orbital phases.
Assuming optically thin conditions, 
the observed variation would require a column density
difference on the order of $10^{11}$~cm$^{-2}$ between the post-transit phases of night~2 and night~1 and, likewise,
during the approximately $1.5$~h post-transit phase of night~2 (cf. Eq.~\ref{eq:tau}).

In Fig.~\ref{fig:hatorus}, we show a hypothetical \ha\ absorption line profile of a homogeneous torus of material
at the planetary distance, covering the entire star. Assuming that the material moves with Keplerian speeds strongly broadens the line.
Due to the circular shape of the star, more material absorbs close to the center of the stellar disk where its vertical is longer and
the material has zero line-of-sight velocity. This gives the profile its almost semi-circular shape, typical of rotational line profiles.
While details such as limb-darkening were not
accounted for and the \ha\ line absorption profile was assumed to be narrow at the instrumental resolution,
this hypothetical line profile is in any case much broader than the observed spectral change between night~1 and 2.
Any heterogeneity in the torus, which would cause time-dependent absorption, would also be subject to Keplerian
motion and, thus, likely produce significant Doppler shifts in the SF.  
At the planetary distance, the Keplerian RV of the material changes at a rate of about $15$~\mbox{m\,s$^{-2}$} or $54$~\mbox{km\,s$^{-1}$\,h$^{-1}$}
during the transit.
We do not see evidence for a significant change in the RV shift of the signal in the SF. Therefore, the signal can hardly
be caused by a change in column density of torus material, moving on a Keplerian orbit
at an orbital separation similar to that of the planetary body. 
If one attributes the observed $-4.4$~\kms\ blueshift of the signal to Keplerian orbital motion during ingress,
an orbit with a radius of about $0.15$~au is required. A change in column density of material farther out in the system,
where Keplerian speeds are lower seems a possible explanation for the observation. Alternatively, different atmospheric geometries may produce narrower absorption profiles.  

\subsection{Interpretation of the \hei\ upper limit}
\label{sec:he_interpretation}

With \wab\ filling $59$\,\% of its Roche lobe volume (Sect.~\ref{sec:Roche}), there remains little room for
an upper atmosphere within it. This is thought to make Roche-lobe overflow the dominant mass-loss mechanism in \wab,
In the Roche-lobe overflow scenario, energy deposited by optical and infrared light
drives the denser, lower atmosphere across the Roche lobe boundary
\citep[e.g.,][]{Koskinen2022, Huang2023}, and the atmospheric density and velocity
at this surface determine the mass-loss rate \citep{Koskinen2022}.
Assuming an isothermal atmosphere for \wab\ with the effective planetary temperature (Table~\ref{tab:props}),
we solved the hypsometric equation \citep[e.g.,][]{Koskinen2022} on the terminator. Assuming a pressure of one bar
at the optical radius, we found that
the atmosphere reaches the Roche lobe at the terminator at a pressure level of
about $10^{-9}$~bar. Therefore, the incoming
XUV radiation is mainly absorbed at or above the Roche lobe, where
it continues to play a
crucial role in the structure of the unbound gas outflow
\citep[e.g.,][]{Huang2023}.

To interpret our upper limits on the \hei\ absorption, we employed the spherically symmetric hydrodynamic
model presented by \citet{Lampon2020}. The latter is a variation of the isothermal
Parker wind approach \citep{Parker1958} with the speed of sound being
held constant throughout the thermosphere. The two main steps in the modeling are
the solution of the hydrodynamical equations and the computation of the population of the
He ($2^3$S) excited state in non-local thermodynamic equilibrium (NLTE), which causes the planetary \hei\ absorption. 
Its free parameters are the temperature, the hydrogen-to-helium number ratio, and the
mass-loss rate; for details, we refer to \citet{Lampon2020}.
Although the actual mass-loss mechanism of \wab\ may be unrelated to the XUV irradiation,
mass continuity combined with our ad hoc treatment of the mass-loss rate  
render the model applicable to simulate the atmosphere from the
thermobase outward, where the \hei\ triplet is produced.

\begin{figure}[]
        \includegraphics[width=0.49\textwidth]{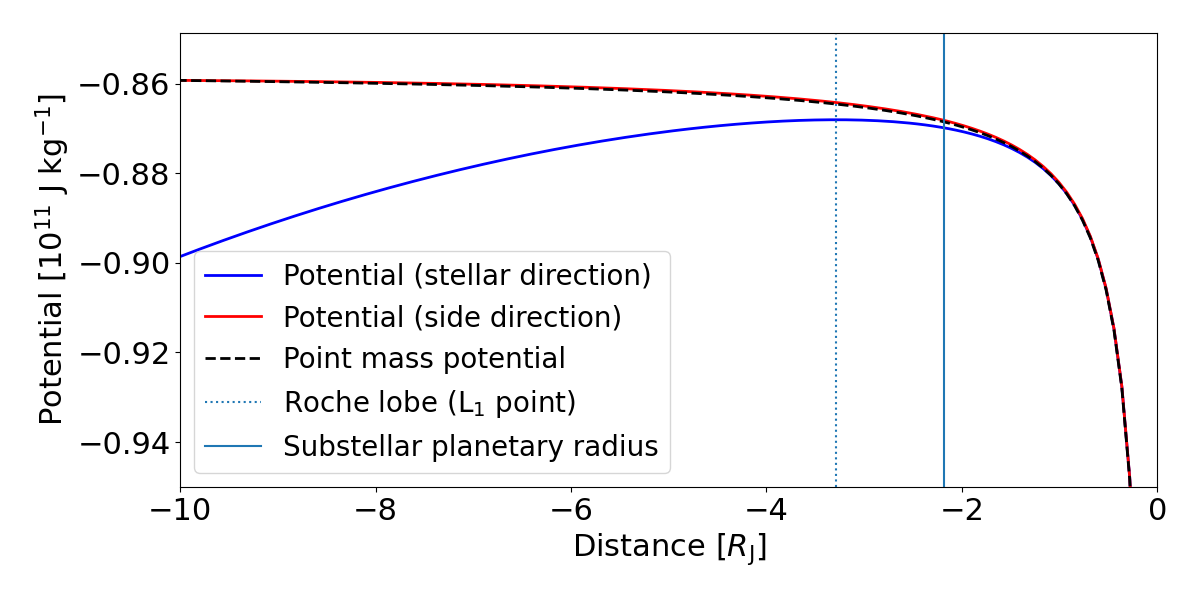}
    \caption{Roche potential around \wab\ evaluated in the orbital plane toward the star (blue)
    and toward the side (red) along with the potential of a point mass (black dashed). The substellar
    radius and the height of the Roche lobe at the substellar point are indicated. The
    acceleration caused by gravitational and centrifugal
    forces in the respective directions is given by the derivative of the potential.    \label{fig:pot}}
\end{figure}

In Fig.~\ref{fig:pot}, we show the Roche potential in the vicinity of \wab\ evaluated in the orbital plane along the
line connecting the star and the planet as well as toward the side. The latter represents an outward radial direction
from the planetary terminator. While the potential to the side only marginally differs from that of an isolated
point mass, the stellar pull and tidal forces strongly affect it in the direction
of the star. To approximate the effect of the modified potential on our modeling, and, in particular, the result of the
lower potential barrier on the mass-loss rate at the substellar point, we set up a terminator and a substellar-point model, differing in
the adopted value of the planetary mass and radius.
Specifically, we used the nominal mass of $1.46$~\mj\ (Table~\ref{tab:props}) to represent the conditions along the terminator
and an effective substellar planetary mass, $M_{\rm p, eff}$, of $0.66$~\mj\ 
and a radius of $2.18$~\rj\
to approximate the conditions at the
substellar point. This effective mass is defined by the condition that
the potential difference between the planetary surface and the Roche lobe at the substellar point (i.e., the
specific energy required to remove material from the planet) matches that of
the Roche potential. A comprehensive treatment of the atmosphere would require the inclusion of the
Roche potential and Coriolis forces in three dimensions (see App.~\ref{sec:rochePot}). Yet, for instance, the study by
\citet{MacLeod2022} showed that spherical planetary wind solutions can still provide good approximations of the
\hei\ transmission spectrum at least in scenarios with comparably weaker Roche effects.

\begin{figure}[]
		\includegraphics[width=0.49\textwidth]{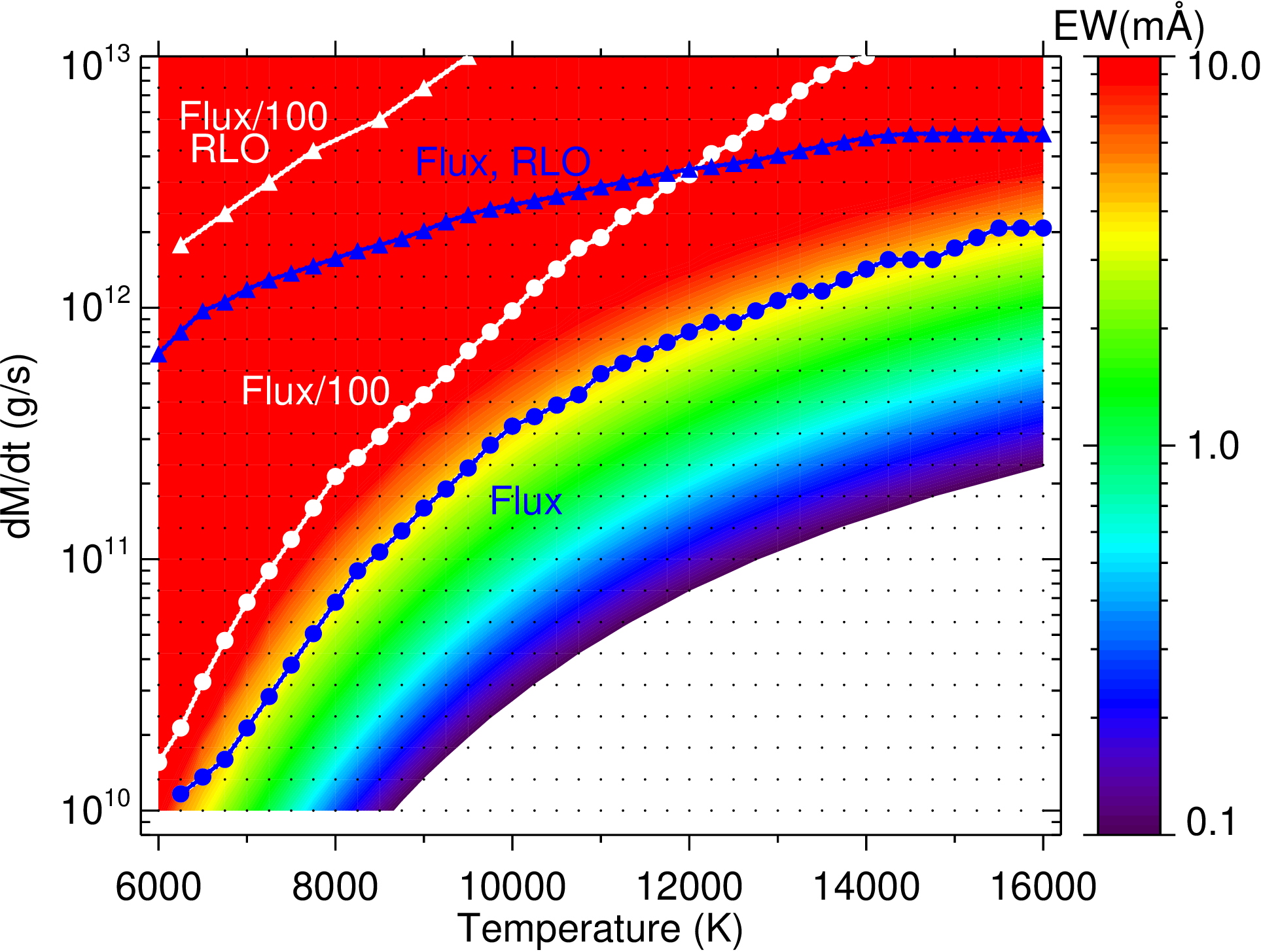}
	\caption{Modeled EW of the \hei\ signal as a function of temperature and mass-loss rate at the terminator
	for the high irradiation scenario. The admissible region is indicated by connected blue circles. Similarly,
	admissible regions are indicated for the moderate irradiation scenario (white, F/100) and the substellar-point
	model in Roche-lobe overflow (RLO, triangles).	\label{fig:ew_vs_tmlr}}
\end{figure}

\begin{figure}[]
		\includegraphics[width=0.49\textwidth]{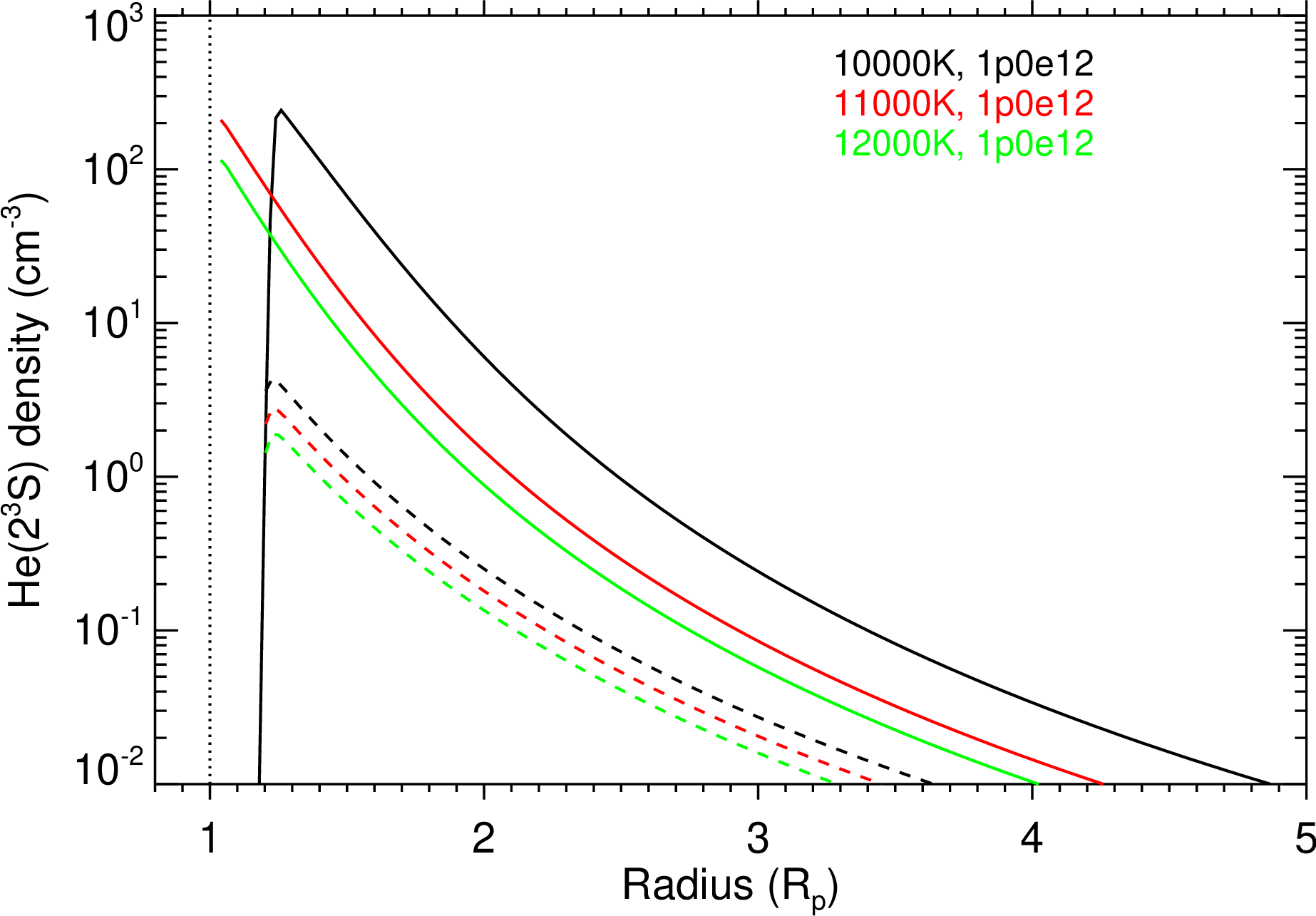}
	\caption{Density of excited helium as a function of distance for the high-irradiation scenario and
	various thermospheric temperatures for the
	terminator (solid) and substellar-point models (dashed); \rp\ is taken to be $1.91$~\rj\ for the plot.
		\label{fig:he_con}}
\end{figure}

Here, we follow a similar approach to \citet{Koskinen2022} and \citet{Huang2023}, who found that
substellar-point models give an upper limit for the mass-loss rate in planets undergoing Roche lobe overflow.As the level of irradiation of \wab\ remains uncertain, we look at two illustrative cases in
some more detail, which we call the high and the moderate irradiation scenario. These are characterized by
XUV fluxes of $318\,000$~\ergcmcms\ at the planetary orbit, corresponding to the nominal upper limit of the stellar XUV flux,
and a factor of 100 less in the moderate scenario. 
We assume a hydrogen-to-helium
number ratio of 99/1, which is in line with the trend found in
other systems \citep{Lampon2021a, Lampon2023}. Finally, a narrow Doppler profile with a FWHM of $0.3$~\AA, located at the
center of the stronger doublet, is
adopted to synthesize the absorption signal. The bulk of such a signal is contained within twice
the FWHM of the smearing-corrected instrumental resolution. Therefore, we adopt an upper limit of $4$~m\AA\ for
the \hei\ signal (Table~\ref{tab:heuls}).
In Fig.~\ref{fig:ew_vs_tmlr}, we show the distribution of the
obtained EWs from our terminator simulations as a function of temperature
and mass-loss rate for the high irradiation scenario and indicate the combinations of mass-loss rate and
temperature consistent with our measurements for both the terminator and substellar-point models.
As we expect the true absorption to be better described
by a weighted average of these models, the borders of these regions outline the extremes of the
expectable mass-loss rate as a function of temperature. 
In both models, the peak of the metastable helium concentration is found within or close
to the Roche lobe (Fig.~\ref{fig:he_con}),
indicating a rather compact thermosphere, which may, therefore, plausibly still be adequately described by a one-dimensional model.
Clearly, the admissible mass-loss rates for the substellar model are significantly higher than those for the terminator model.

\begin{table}[]
	\caption{Gravitational potential in Jovian units, XUV irradiation at planetary separation (5--920~\AA, see Sect.~\ref{sec:xmm}),
	and derived thermospheric temperature range for relevant systems from \citet{Lampon2023}.
	 \label{tab:heicomp}}
	 \begin{tabular}[h]{l c c c} 
	 	\hline\hline
        \noalign{\smallskip}
	 	System & $\Phi$ & $F_{\rm XUV}$ & $T$ \\
	 		& [$\Phi_{\rm Jup}$] & [\ergcmcms] & [K] \\
	        \noalign{\smallskip}
 		\hline
        \noalign{\smallskip}
	 	\wab\ & 0.77  & $\lesssim 318\,000$ & ... \\
		HD 189733\,b & 0.94 & 25\,400 & $12\,700 \pm 900$ \\
		HD 209458\,b & 0.50 & 2\,400 & $7600 \pm 500$ \\
 		WASP--76\,b & 0.48 & 145\,600 & $11\,500 \pm 5500$ \\
 		\hpb\ & 0.33 & 417\,400 & $12\,400 \pm 2900$ \\
        \noalign{\smallskip}
 		\hline
	 \end{tabular}
\end{table}

To narrow down the likely location of \wab\ in the diagram of Fig.~\ref{fig:ew_vs_tmlr},
plausible values for the thermospheric temperature and the mass-loss rate are needed.
To that end, we put the planet into the context of the sample study
by \citet{Lampon2023}.
In the high-flux scenario, the XUV irradiation level of \wab\ is situated
between those of WASP--76\,b and \hpb\ (Table~\ref{tab:heicomp}), which puts \wab\ in the recombination-limited escape
regime, characterized by a largely ionized thermosphere \citep{Lampon2021b}.
For WASP--76\,b and \hpb\ temperatures of about $12\,000$~K and mass-loss rates
in the range of $10^{12}$~\gs\ to $10^{13}$~\gs\ were found.
At 12\,000~K, we find limits of $8\times 10^{11}$~\gs\ and $4\times 10^{12}$~\gs\ for the upper limit of the
mass-loss rate. The conservative upper limit is, thus, given by the higher value, while lower values may be expected
depending on the contribution of the terminator region to the atmospheric absorption.
A key difference between
\wab\ and both WASP--76\,b and \hpb\ lies in the higher gravitational potential of \wab, which may contribute
to the conspicuous lack of a \hei\ signal in \wab.

In the moderate irradiation scenario, the irradiating XUV flux of \wab\ becomes comparable to that
of HD~209458\,b. For such moderately irradiated planets, thermospheric temperature correlates with
gravitational potential \citep{Salz2016b, Lampon2023}.
The latter is situated between those of
HD~209458\,b and HD~189733\,b, which is, however, more strongly irradiated (Table~\ref{tab:heicomp}).
Using these as
bounding cases yields a temperature estimate in the range of $7500$~K to $12\,500$~K.
Adopting a plausible temperature estimate of $10\,000$~K, we find upper limits of $10^{12}$~\gs\
for the terminator model to be consistent with our non-detection of \hei\ absorption.
In the substellar-point model, the lower gravitational potential well leads to a faster
outflow of gas, which produces stronger adiabatic cooling. Under such conditions, we expect
a lower value of around $7000$~K for the maximum temperature, which yields a mass-loss rate
of about $4\times 10^{12}$~\gs.

In particular, in the case of the moderate irradiation scenario, the excited \hei\ atmosphere
maybe be rather extended at the terminator so that geometrical effects caused by the partial
overlap with the stellar disk and of course the
limitations of a one-dimensional representation can become relevant. 
However, such a large extent may be unrealistic because the flow is confined by other factors.
Using a two-dimensional model, \citet{Dwivedi2019}
found that strong stellar winds induce a lateral ionopause boundary for which they estimated
altitudes between two and four planetary radii at the terminator
and no significant confinement at the substellar point. To evaluate the effect of such a confinement in our results,
we took into account the ionopause by truncating the atmosphere as demonstrated by \citet{Lampon2023}.
We found that the mass-loss rates are similar or lower at any given temperature, so that
our upper limits on the mass-loss rates remain valid even in a strong stellar wind environment.

While we cannot rule out either irradiation scenario, the combination of the derived upper limit on the \hei\ EW
in transmission and our models for atmospheric absorption slightly favors the
moderate irradiation scenario as outlined above, with
mass-loss rates of $\lesssim 4\times 10^{12}$~\gs\ for the substellar-point model, which is
consistent with the results of \citet{Dwivedi2019}, who estimated a mass-loss rate of $10^{12}$~\gs\ based on
\ion{Mg}{ii} absorption from \wab.
We caution, however, that there are many assumptions in our analysis by necessity. 

\subsection{The lack of \nad\ absorption}

The cores of the \nad\ doublet lines belong to the most sensitive atomic tracers of the atmosphere \citep[e.g.,][]{Brown2001}.
In our analysis of \wab, we only find upper limits for their strength. 
As shown in Sect.~\ref{sec:he_interpretation}, the gravitational potential of \wab\ at the terminator
is almost that of a point mass. 
Using \texttt{petitRadtrans} \citep{Molliere2019}, we calculated synthetic transmission spectra for an isothermal hydrogen-helium
plus sodium atmosphere with a mean molecular weight of $2.3$ and sodium mass fractions of up to $10^{-3}$ (i.e., about
30 time the solar value). These simulations predict EWs of $\le 1$~m\AA\ for the cores of the \nad\ lines in the transmission spectrum
if the atmosphere is truncated at $10^{-9}$~bar, where it reaches the Roche lobe (Sect.~\ref{sec:he_interpretation}).
Extending it beyond that up to $10^{-12}$~bar, the EWs
become about $20$\,\% larger. Any of these values are consistent with our upper limit of about $2$~m\AA\ (Table~\ref{tab:alluls}), together with
the presumed presence of a cloud deck on \wab\ \citep{Wakeford2017}, which would tend to further decrease
the absorption signal.

\begin{figure}[]
        \includegraphics[width=0.49\textwidth]{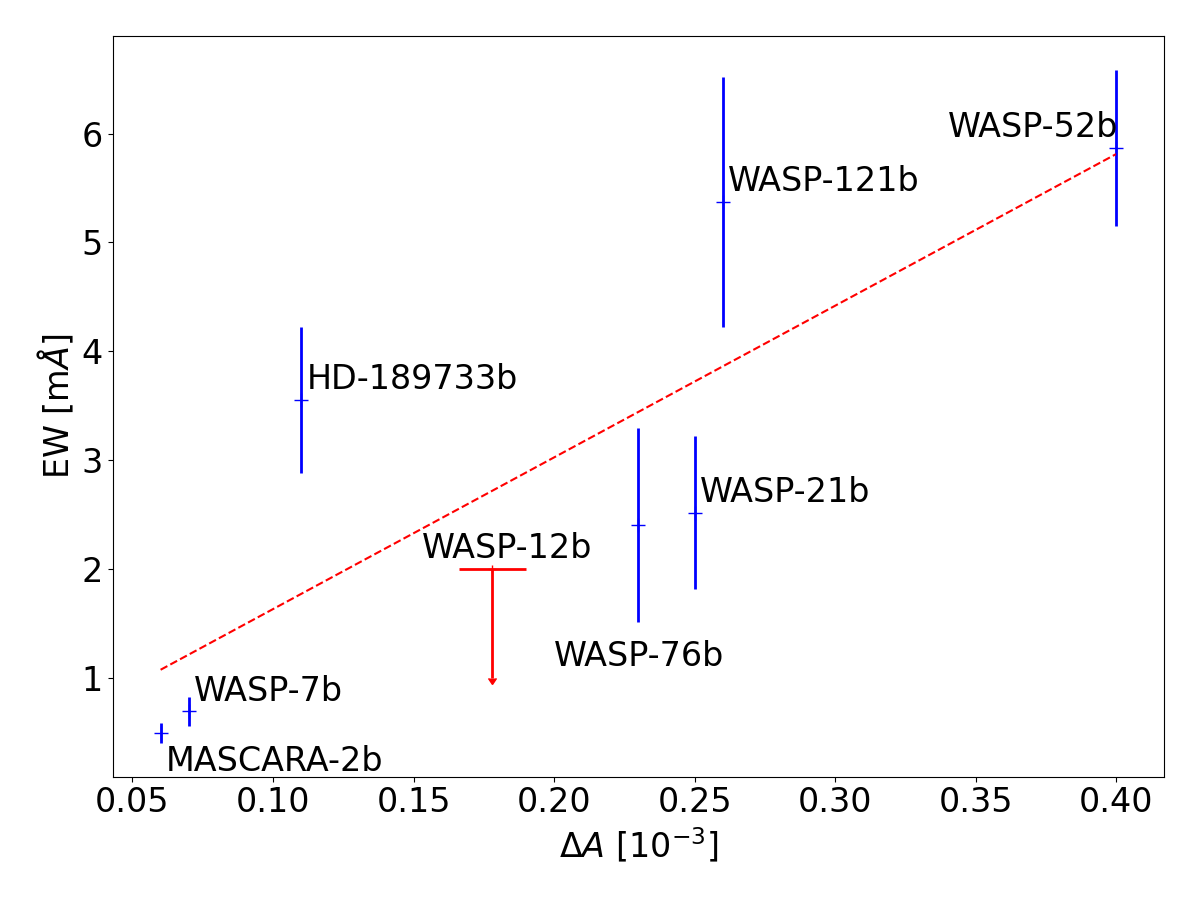}
    \caption{Absorption EW in the \nad$_2$ line as a function of the fractional atmospheric coverage fraction per scale height, $\Delta A$,
    for a sample of planets along with \wab. 
    \citep[data adopted from][]{Wyttenbach2015, Seidel2019, CasasayasBarris2019, Chen2020, Chen2020a, Cabot2020, Langeveld2022, Rahmati2022}. 
    \label{fig:deltaA}}
\end{figure}

In Fig.~\ref{fig:deltaA}, we put our upper limit into the context of previous detections. In particular, we show the EW of the
core of the \nad$_2$ line as a function of the fractional atmospheric coverage fraction per scale height \citep{Rahmati2022}
\begin{equation}
    \Delta A = \frac{2 \mbox{\rp} H}{\mbox{\rs}^2} ,
\end{equation}
where $H$ is the atmospheric scale height, for which we obtained a value of $880$~km at the terminator of \wab.
The cited measurements suggest a trend, which is, however, not (yet) overly significant
(Person's correlation coefficient has a value of $0.82$ with a $p$-value of $2.4$\,\%).
While the \nad\ absorption in \wab\ may still be small in the context of the existing sample,
an unusually high value can be ruled out.

\section{Conclusion}
We analyzed two high-resolution spectral time series of \wab\ obtained with \carm\ about one year apart.
Both series cover about six~hours and are approximately centered
on the optical transit. Our data contain the \hei\ triplet lines with high resolution for the first time and simultaneously provide continuous
coverage of the \ha\ lines. 
The spectrum of \wa\ shows strong interstellar features in the \nad\ lines, the strengths of which we found to be consistent with previous
measurements along a close-by line of sight. 

The in-transit transmission spectrum of \wab\ was obtained with the help of the \sys\ algorithm and the
technique of signal protection. We found \sys\ to be
well suited to carry out atomic line transmission spectroscopy in planets showing sufficiently large orbital
radial velocity variation.
A comparison with a conventional approach showed comparable results. 
The obtained transmission spectrum is devoid of all tested atomic features in the PF to within our sensitivity limits.
This result is consistent with the non-detections by \citet{Sing2013}, but based on considerably higher spectral resolution, which
specifically adds sensitivity to the line cores.
The upper limit on the planetary \nad\ features is consistent with predictions by isothermal atmospheric models and
previous measurements in other similar systems.
The cores of the \ha\ and the \hei\ lines of the system show variations between night~1 and 2 in the SF, primarily after
the optical transit. The characteristics of this variation are
difficult to explain by a heterogeneous torus of material at the planetary orbital separation or by stellar activity.
We speculate that a change in column density of \ha\ absorbing material farther out in the system might cause the observed
behavior. This does not exclude
a large structure fed by planetary material similar to the one recently observed in the \hp\ system \citep{Zhang2023},
but only that the observed changes between night~1 and 2 are unlikely to be caused by it.

Our analysis of \xmm\ observations of \wa\ yields a marginal detection of X-rays from the system, which we interpreted as
an upper limit. Combining this upper limit with the absence of a \hei\ absorption signal and models of the upper planetary atmosphere
slightly favors a moderate irradiation level for the planet, with a thermospheric temperature of $\lesssim 12\,000$~K, and
a conservative upper limit of $\lesssim 4\times 10^{12}$~\gs\ on the mass-loss rate,
consistent with previous estimates based on \ion{Mg}{ii} absorption \citep{Dwivedi2019}.
We caution, however, that this result
relies on a number of assumptions. Overall, our results provide no evidence against the hypothesis that \wa\ is, indeed, a very
inactive, slowly rotating star. Although the state of \wab's elusive atmosphere remains enigmatic, we put more stringent limits
on it.

\begin{acknowledgements}
CARMENES is an instrument at the Centro Astron\'omico Hispano en Andaluc\'ia (CAHA) at Calar Alto (Almer\'{\i}a, Spain),
operated jointly by the Junta de Andaluc\'ia and the Instituto de Astrof\'isica de Andaluc\'ia (CSIC).
The authors wish to express their sincere thanks to all members of the Calar Alto staff for their expert support of the instrument and telescope operation.
  CARMENES was funded by the Max-Planck-Gesellschaft (MPG), 
  the Consejo Superior de Investigaciones Cient\'{\i}ficas (CSIC),
  the Ministerio de Econom\'ia y Competitividad (MINECO) and the European Regional Development Fund (ERDF) through projects FICTS-2011-02, ICTS-2017-07-CAHA-4, and CAHA16-CE-3978, 
  and the members of the CARMENES Consortium 
  (Max-Planck-Institut f\"ur Astronomie,
  Instituto de Astrof\'{\i}sica de Andaluc\'{i}a,
  Landessternwarte K\"onigstuhl,
  Institut de Ci\`encies de l'Espai,
  Institut f\"ur Astrophysik G\"ottingen,
  Universidad Complutense de Madrid,
  Th\"uringer Landessternwarte Tautenburg,
  Instituto de Astrof\'{\i}sica de Canarias,
  Hamburger Sternwarte,
  Centro de Astrobiolog\'{\i}a and
  Centro Astron\'omico Hispano-Alem\'an), 
  with additional contributions by the MINECO, 
  the Deutsche Forschungsgemeinschaft (DFG) through the Major Research Instrumentation Programme and Research Unit FOR2544 ``Blue Planets around Red Stars'', 
  the Klaus Tschira Stiftung, 
  the states of Baden-W\"urttemberg and Niedersachsen, 
  and by the Junta de Andaluc\'{\i}a.
  We acknowledge financial support from the Agencia Estatal de Investigaci\'on (AEI/10.13039/501100011033) of the Ministerio de Ciencia e Innovaci\'on and the ERDF ``A way of making Europe'' through projects 
  PID2021-125627OB-C31,	PID2019-110689RB-I00,	  PID2019-109522GB-C5[1:4],	  and the Centre of Excellence ``Severo Ochoa'' and ``Mar\'ia de Maeztu'' awards to the Instituto de Astrof\'isica de Canarias (CEX2019-000920-S), Instituto de Astrof\'isica de Andaluc\'ia (CEX2021-001131-S) and Institut de Ci\`encies de l'Espai (CEX2020-001058-M).
This work was also funded by the Generalitat de Catalunya/CERCA programme, 
the LMU Munich Astrophysics Fraunhofer and Schwarzschild Postdoctoral Fellowship,
and the DFG through projects 390783311 (under Germany's Excellence Strategy EXC 2094) and 314665159, 
and priority program SPP 1992 ``Exploring the Diversity of Extrasolar Planets'' (CZ 222/5-1, CZ 222/3-1),
\end{acknowledgements}

\bibliographystyle{aa}
\bibliography{wasp12_carm}

\newpage
\clearpage
\begin{appendix}

\section{Observation log}
\label{sec:obslog}

Observation log for our \carm\ observations of \wab. The numbers refer to
the VIS channel spectra. The values for the corresponding NIR channel observations
differ only slightly.

\begin{table}[]
\caption{Observation log of night~1$^a$. 
\label{tab:obslog1}}
\centering
\begin{tabular}{c c c c c}
\hline\hline
\noalign{\smallskip}
\# & BJD$_{\rm TDB}$ & Phase & Overlap & Airmass \\
\noalign{\smallskip}
\hline
\noalign{\smallskip}
 1 & $0.34776$ & $-0.0996$ & $0.00$ & 1.96 \\
 2 & $0.36178$ & $-0.0867$ & $0.00$ & 1.77 \\
 3 & $0.37285$ & $-0.0766$ & $0.00$ & 1.64 \\
 4 & $0.38482$ & $-0.0656$ & $0.00$ & 1.53 \\
 5 & $0.39619$ & $-0.0552$ & $0.07$ & 1.43 \\
 6 & $0.40942$ & $-0.0431$ & $1.00$ & 1.35 \\
 7 & $0.42015$ & $-0.0333$ & $1.00$ & 1.28 \\
 8 & $0.43145$ & $-0.0229$ & $1.00$ & 1.23 \\
 9 & $0.44341$ & $-0.0119$ & $1.00$ & 1.18 \\
10 & $0.45357$ & $-0.0026$ & $1.00$ & 1.15 \\
11 & $0.46746$ & $+0.0101$ & $1.00$ & 1.11 \\
12 & $0.47958$ & $+0.0212$ & $1.00$ & 1.08 \\
13 & $0.48986$ & $+0.0306$ & $1.00$ & 1.06 \\
14 & $0.50379$ & $+0.0434$ & $1.00$ & 1.04 \\
15 & $0.51460$ & $+0.0533$ & $0.21$ & 1.02 \\
16 & $0.52532$ & $+0.0631$ & $0.00$ & 1.02 \\
17 & $0.53705$ & $+0.0739$ & $0.00$ & 1.01 \\
18 & $0.54807$ & $+0.0840$ & $0.00$ & 1.01 \\
19 & $0.55856$ & $+0.0936$ & $0.00$ & 1.01 \\
20 & $0.57072$ & $+0.1047$ & $0.00$ & 1.02 \\
21 & $0.58175$ & $+0.1148$ & $0.00$ & 1.02 \\
22 & $0.59229$ & $+0.1245$ & $0.00$ & 1.04 \\
\noalign{\smallskip}
\hline
\end{tabular}
\tablefoot{$^a$ The columns give a running number,
the time at middle of exposure BJD$_{\rm TDB} - 2458836$, the orbital
phase according to the quadratic ephemeris by \citet{Turner2021},
the time-averaged geometric overlap of the planetary and stellar disks during the exposure, and the airmass.}
\end{table}

\begin{table}[]
\caption{Same as Table~\ref{tab:obslog1} for night~2 but with a time offset of 2459171.
\label{tab:obslog2}}
\centering
\begin{tabular}{c c c c c}
\hline\hline
\noalign{\smallskip}
\# & BJD$_{\rm TDB}$ & Phase & Overlap & Airmass \\
\noalign{\smallskip}
\hline
\noalign{\smallskip}
 1 & $0.42713$ & $-0.0866$ & $0.00$ & 2.00 \\
 2 & $0.43917$ & $-0.0756$ & $0.00$ & 1.82 \\
 3 & $0.45085$ & $-0.0649$ & $0.00$ & 1.67 \\
 4 & $0.46298$ & $-0.0538$ & $0.17$ & 1.55 \\
 5 & $0.47534$ & $-0.0425$ & $1.00$ & 1.45 \\
 6 & $0.48757$ & $-0.0313$ & $1.00$ & 1.36 \\
 7 & $0.49852$ & $-0.0212$ & $1.00$ & 1.30 \\
 8 & $0.51396$ & $-0.0071$ & $1.00$ & 1.23 \\
 9 & $0.52480$ & $+0.0028$ & $1.00$ & 1.18 \\
10 & $0.53840$ & $+0.0153$ & $1.00$ & 1.14 \\
11 & $0.54909$ & $+0.0251$ & $1.00$ & 1.11 \\
12 & $0.56133$ & $+0.0363$ & $1.00$ & 1.08 \\
13 & $0.57182$ & $+0.0459$ & $0.88$ & 1.06 \\
14 & $0.58292$ & $+0.0561$ & $0.02$ & 1.04 \\
15 & $0.59524$ & $+0.0674$ & $0.00$ & 1.03 \\
16 & $0.60621$ & $+0.0774$ & $0.00$ & 1.02 \\
17 & $0.61737$ & $+0.0877$ & $0.00$ & 1.01 \\
18 & $0.62949$ & $+0.0988$ & $0.00$ & 1.01 \\
19 & $0.63995$ & $+0.1084$ & $0.00$ & 1.01 \\
20 & $0.65087$ & $+0.1184$ & $0.00$ & 1.02 \\
21 & $0.66352$ & $+0.1299$ & $0.00$ & 1.03 \\
\noalign{\smallskip}
\hline
\end{tabular}
\end{table}

\section{Convolution and multiplication}
\label{sec:convolution_distributive}
Multiplication and convolution cannot generally be exchanged, but the error 
incurred by it may be acceptable nonetheless.
Let us assume there are two functions $F(\lambda)$ and $g(\lambda)$ and a profile function $\mathcal{P}(\lambda)$. We
know the results of $\mathcal{P}(\lambda) * \left( F(\lambda)\, g(\lambda) \right)$ and
$\mathcal{P}(\lambda) * F(\lambda)$ and need to calculate $\mathcal{P}(\lambda) * g(\lambda)$.
Potential other methods notwithstanding, this would be a simple matter
if we could write
$\mathcal{P}(\lambda) * \left( F(\lambda)\, g(\lambda) \right) = (\mathcal{P}(\lambda) * F(\lambda)) \times (\mathcal{P}(\lambda) * g(\lambda))$,
that is, if convolution were (left) distributive over multiplication, which is, unfortunately, not the case.
To estimate the magnitude of the deviation 
\begin{equation}
  	 \delta_{\rm g}(\lambda) =
  	 \frac{ \mathcal{P}(\lambda) * \left( F(\lambda)\, g(\lambda) \right) }{\mathcal{P}(\lambda) * F(\lambda)} - \mathcal{P}(\lambda) * g(\lambda)
	\label{eq:conv_approx}
\end{equation}
incurred by the assumption, we need to specify the functions.

\begin{figure}[]
			\includegraphics[width=0.49\textwidth]{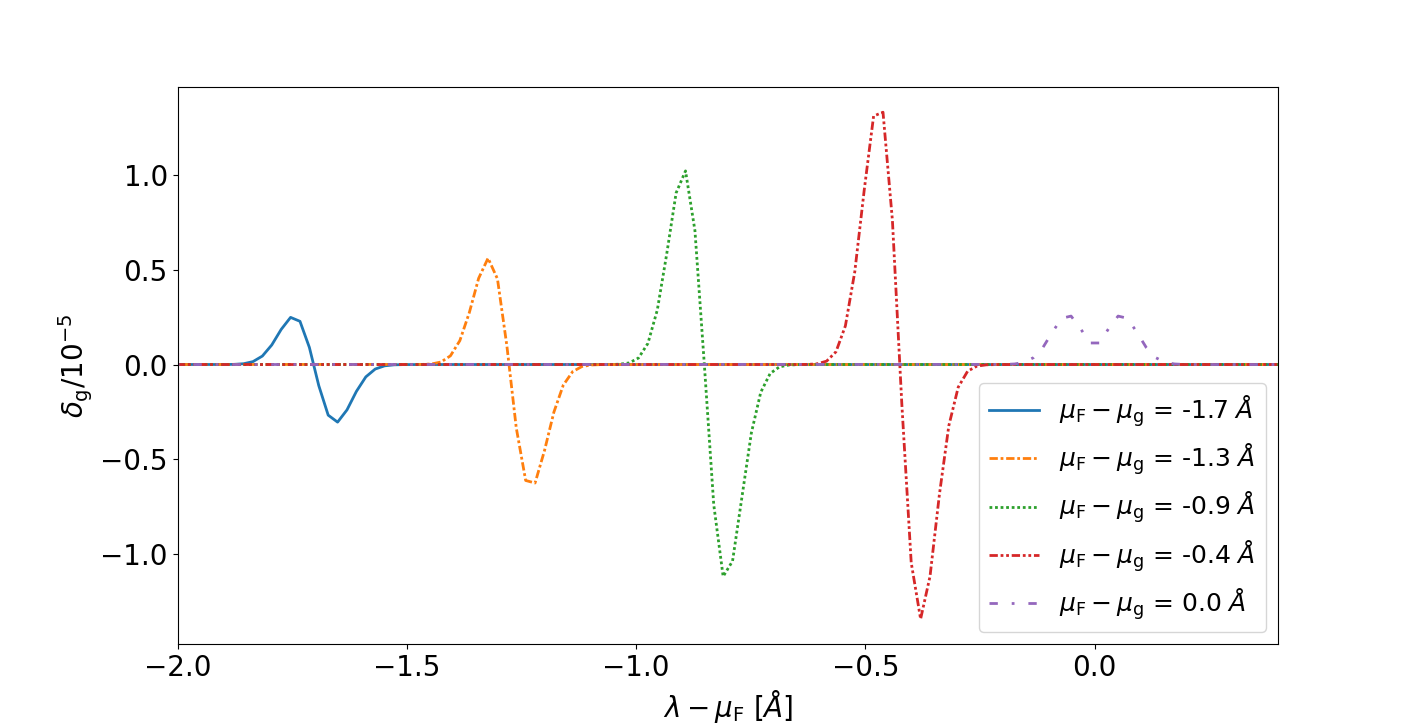}
	\caption{Error as defined by Eq.~\ref{eq:conv_approx} incurred by assuming distributivity of convolution over multiplication.   
	\label{fig:deltag}}
\end{figure}

To get a handle on our problem, we assume that all functions are Gaussian. With
$\mathcal{N}(\mu, \sigma)$ denoting the density of the normal distribution with mean $\mu$ and standard deviation $\sigma$,
we specify
\begin{align}
	\mathcal{P(\lambda)} &= \mathcal{N}(0, \sigma_0), \nonumber \\
	F(\lambda) &= 1 - d_{\rm F}\sqrt{2\pi\sigma_{\rm F}^2} \mathcal{N}(\mu_{\rm F}, \sigma_{\rm F}), \; \mbox{and} \nonumber \\
	g(\lambda) &= 1 - d_{\rm g}\sqrt{2\pi\sigma_{\rm g}^2} \mathcal{N}(\mu_{\rm g}, \sigma_{\rm g}) .
\end{align}
Here, $d_{\rm F}$ and $d_{\rm g}$ parametrize the depths of the Gaussians. 
With this choice, an analytic though rather convoluted form for the expression in Eq.~\ref{eq:conv_approx} can be derived using, for instance,
the computer algebra system \texttt{maxima} \citep{maxima}. 
In our experiment, we
adopt the specific values $\sigma_0 = 0.032$\,\AA, $d_{\rm F} = 0.9$, $\sigma_{\rm F}=0.85$\,\AA, $\mu_{\rm F} = 6564$\,\AA,
$d_{\rm g} = 10^{-3}$, and $\sigma_{\rm g}=\sigma_0$. The standard deviation of the profile function, $\sigma_0$,
is compatible with the instrumental resolution around
the \ha\ line, which is represented by the function $F(\lambda)$, while $g(\lambda)$ models a hypothetical
planetary absorption signal, for which we assume a depth of $10^{-3}$. In Fig.~\ref{fig:deltag}, we show the
deviation caused by assuming distributivity of convolution over multiplication (Eq.~\ref{eq:conv_approx}) for various placements of the
absorption signal within the trough. In all tested cases, the difference is about two orders of magnitude smaller than
the absorption signal, which we intend to recover. Therefore, at least in this case, we consider the approximation justifiable. 

\section{Planetary equilibrium temperature}
\label{sec:teq}

A formula for the planetary equilibrium temperature, \tc{eq}, valid for close-in planets, where the usual assumption of a small ratio between
the stellar radius, $\rs$, and the semi-major axis, $a$, is no longer accurate is given by 
\begin{equation}
	\mbox{\tc{eq}} = \mbox{\tc{eff}} \left(\frac{1}{2} \left(1-\sqrt{1- \left(\frac{\rs}{a}\right)^2 }\right)(1-A_{\rm B}) \right)^{\frac{1}{4}} ,
	\label{eq:teq}
\end{equation}
where \tc{eff} is the stellar effective temperature and $A_{\rm B}$ the planetary Bond albedo.
As shown below, Eq.~\ref{eq:teq} still requires
$a \gg \mbox{\rp}$. 
For $\rs \ll a$, Eq.~\ref{eq:teq} readily reduces to the usual expression by a Taylor expansion of the square root
\begin{equation}
	\mbox{\tc{eq}} \approx \mbox{\tc{eff}} \left( \left(\frac{\rs}{2 a}\right)^2  (1-A_{\rm B}) \right)^{\frac{1}{4}} .
\end{equation}

A derivation of Eq.~\ref{eq:teq} can be formulated on the grounds of geometrical arguments.  
The planetary equilibrium temperature is defined by the condition that the rate of radiative energy absorbed by the (spherical)
planet, $\dot{E}_{\rm p, abs}$, from the star
is balanced by the energy homogeneously re-radiated from its surface into space via thermal back-body emission,
$L_{\rm p, out} = 4\pi R_{\rm p}^2 \sigma T_{\rm eq}^4$. If the planet has non-zero bond albedo,
only a fraction of the available incoming radiation,
$L_{\rm p, in}$, is absorbed so that $\dot{E}_{\rm p, abs} = (1-A_{\rm B}) L_{\rm p, in}$.

In the hypothetical case of a ``star'' covering the entire sky ($\Omega_{\star} = 4\pi$), radiative equilibrium demands
\begin{align}
   \dot{E}_{{\rm p, abs}, \Omega_{\star} = 4\pi} &= L_{{\rm p}, \Omega_{\star} = 4\pi} \\
   4\pi \sigma R_{\rm p}^2 T_{\rm eff}^4 (1-A_{\rm B}) &= 4\pi \sigma R_{\rm p}^2 T_{\rm eq}^4.
\end{align}
and, thus,
\begin{equation}
    \left(\frac{T_{\rm eq}}{T_{\rm eff}} \right)^4 _{\Omega_{\star} = 4\pi} = 1-A_{\rm B} .
\end{equation}
For a perfect black body ($A_{\rm B} = 0$), the effective and equilibrium temperatures are equal. 
If the solid angle subtended by the star is reduced, the amount of radiation absorbed by the spherical
planet decreases in proportion so that
\begin{equation}
   \dot{E}_{\rm p, abs}(\Omega_{\star}) = \dot{E}_{{\rm p, abs}, \Omega_{\star} = 4\pi} \times \frac{\Omega_{\star}}{4\pi}
    \overset{!}{=} L_{\rm p, out},
\end{equation}
which yields
\begin{equation}
    \left( \frac{T_{\rm eq}}{T_{\rm eff}} \right)^4 = \frac{\Omega_{\star}}{4\pi} (1-A_{\rm B})   .
    \label{eq:condition}
\end{equation}
Now, the solid angle subtended by a spherical star with radius $\rs$,
viewed from a point at a distance $a$ from its center,
is that of a cone with half opening angle $\theta = \arcsin\left(\rs\,a^{-1} \right)$, which
is given by
\begin{align}
    \Omega_{\star}(a, \rs) &=  2\pi\left(1-\cos\left(\arcsin\left(\frac{\rs}{a}\right)\right) \right) \nonumber \\
    &= 2\pi\left(1 - \sqrt{1-\left( \frac{\rs}{a}\right)^2 } \right) .
    \label{eq:cone_solid_angle}
\end{align}  
Substituting Eq.~\ref{eq:cone_solid_angle} into Eq.~\ref{eq:condition} yields Eq.~\ref{eq:teq}.
In the latter substitution, we implicitly assumed that Eq.~\ref{eq:cone_solid_angle} holds for
every point on the planetary surface, which implies $\mbox{\rp} \ll a$.

\section{The Roche potential}
\label{sec:rochePot}

For the star placed at the origin of a right-handed coordinate system, corotating with the
orbital motion of a planet located at $(a,0,0)$,
the Roche potential, $\phi_{\rm R}$, is given by \citep[e.g.,][]{Hilditch2001}
\begin{equation}
    \phi_{\rm R}(M_{\rm tot},a,x',y',z',q) = -\frac{GM_{\rm tot}}{2a} \phi_{\rm n}(x', y', z', q) ,
    \label{eq:rochePot}
\end{equation}
where the total mass, $M_{\rm tot}$, is $M_{\star} + \mbox{\mp}$, the mass ratio, $q$, is given by $\mbox{\mp}\,M_{\star}^{-1}$, $G$
is the gravitational constant,
and the dimensionless coordinates $x'$, $y'$, and $z'$ are obtained through scaling by the orbital separation
\begin{equation}
    x', y', z' = \frac{x}{a}, \frac{y}{a}, \frac{z}{a} .
\end{equation}
The dimensionless distances from the star and planet, $r'_{\star}$ and $r'_{\rm p}$, are given by
\begin{align}
    r'_{\star} &= \sqrt{x'^2 + y'^2 + z'^2} \nonumber \\
    r'_{\rm p} &= \sqrt{(x'-1)^2 + y'^2 + z'^2} .
\end{align}
Finally, the dimensionless Roche potential $\phi_{\rm n}$ has the form
\begin{align}
     \phi_{\rm n}(x', y', z', q) =& \frac{2}{(1+q)r_{\star}'} + \frac{2q}{(1+q)r_{\rm p}'} + \nonumber \\
     & \left(x' - \frac{q}{1+q} \right)^2 + y'^2 \; ,
\end{align}
which completes the specification of Eq.~\ref{eq:rochePot}. As the coordinate system rotates, moving matter is additionally subject to
Coriolis forces, which cannot be represented by a scalar potential.
For small mass ratios $q \ll 1$ and small planets, the Roche potential in the vicinity of the planet ($r'_{\rm p} << 1$) reduces to
the usual expression of a point mass
\begin{equation}
    \phi_{\rm R, p} \approx -G \frac{\mbox{\mp}}{r'_{\rm p} a} = -G \frac{\mbox{\mp}}{r_{\rm p}} .
\end{equation}

\end{appendix}

\end{document}